\documentclass[fleqn,usenatbib]{mnras}

\usepackage{newtxtext,newtxmath}

\usepackage[T1]{fontenc}
\usepackage{ae,aecompl}
\usepackage{fontawesome}

\usepackage[singlelinecheck=off,justification=justified]{subcaption}
\usepackage{graphicx}	
\usepackage{amsmath}	
\usepackage{amssymb}	
\usepackage{stmaryrd}
\usepackage[usestackEOL]{stackengine}
\stackMath

\usepackage{bm}
\usepackage[mathscr]{euscript}
\usepackage{enumitem}

\newcommand{\betahat}{\hat{\bm \beta}}

\newcommand{\diff}{\textrm{{d}}}
\newcommand{\gammahat}{\hat{\bm n}}
\newcommand{\Ker}[3]{_{#1}\mathcal{K}^m_{{#2}{#3}}}
\newcommand{\C}{\bm{\mathbb{C}}}  

\title[Footprints of Doppler and Aberration Effects in CMB Experiments]{Footprints of Doppler and Aberration Effects in CMB Experiments: Statistical and Cosmological Implications}

\author[S. Yasini \& E. Pierpaoli]{
Siavash Yasini,$^{1}$\thanks{yasini@usc.edu}
Elena Pierpaoli,$^{2}$\thanks{pierpaol@usc.edu}
\\
$^{1}$Physics \& Astronomy Department, University of Southern California, Los Angeles, California,  90089-0484 \\
}

\date{Accepted XXX. Received YYY; in original form ZZZ}

\pubyear{2020}

\begin{document}
\label{firstpage}
\pagerange{\pageref{firstpage}--\pageref{lastpage}}
\maketitle

\begin{abstract}

In the frame of the Solar System, the Doppler and aberration effects cause distortions in the form of mode couplings in the cosmic microwave background (CMB) temperature and polarization power spectra and hence impose biases on the statistics derived by the moving observer. 
We explore several aspects of such biases and pay close attention to their effects on CMB polarization which previously have not been examined in detail. A potentially important bias that we introduce here is \emph{boost variance}---an additional term in cosmic variance, induced by the observer's motion. Although this additional term is negligible for whole-sky experiments, in partial-sky experiments it can reach 10\% (temperature) -- 20\% (polarization) of the standard cosmic variance ($\sigma$). 
Furthermore, we investigate the significance of motion-induced \emph{power} and \emph{parity} asymmetries in TT, EE, and TE as well as potential biases induced in cosmological parameter estimation performed with whole-sky TTTEEE. Using Planck-like simulations, we find that our local motion induces $\sim1-2 \%$ hemispherical asymmetry in a wide range of angular scales in the CMB temperature and polarization power spectra; however, it does not imply any significant amount of parity asymmetry or shift in cosmological parameters. Finally, we examine the prospects of measuring the velocity of the Solar System w.r.t. the CMB with future experiments via the mode coupling induced by the Doppler and aberration effects. Using the CMB TT, EE, and TE power spectra up to $\ell=4000$, SO and CMB-S4 can make a dipole-independent measurement of our local velocity respectively at $8.5\sigma$ and $20\sigma$.
\end{abstract}



\section{Introduction}\label{sec:intro}

The motion of the Solar System with respect to the cosmic microwave background (CMB) creates mode couplings in the observed temperature and polarization anisotropies. These mode couplings can be interpreted as leakage of nearby harmonic multipoles into each other in the moving frame, inducing a scale-dependent change in the estimated power spectrum \citep{Challinor2002}. The leakage is a result of the relativistic Doppler and aberration effects (Lorentz boost) caused by the motion of the observer (us) in the rest frame of the CMB. The largest component of this effect is the well-known CMB dipole moment---which is the rest frame monopole leaking into the observed dipole in the moving frame \citep{Yasini:2016dnd, Kamionkowski1997a, Yasini:2016pby}---but the leakage exists among other multipoles as well. The mode coupling arising from the motion-induced leakage is usually disregarded, with the exception of the dipole, but it has been shown that depending on the geometry and sky/frequency coverage of the specific CMB experiment it can have non-trivial consequences and can potentially lead to biases in CMB statistics \citep{Yasini:2017jqg, Chluba2011, Dai2014, Jeong:2013sxy}.

In the half portion of the sky towards the direction of motion, the primary effects of the boost on the power spectrum are (i) an overall increase in total power, and (ii) a decrease in the size of angular fluctuations\footnote{This can be interpreted as power leaking to higher multipoles in the direction of motion.} \citep{Yasini:2017jqg}. The opposite holds in the antipodal half of the sky. These changes (i \& ii) naturally lead to an asymmetry in the observed CMB power spectrum and any other statistics inferred from opposite halves of the sky. In this paper we examine several statistics that could be potentially affected by the boost, and assess the amount of motion-induced bias in them using realistic simulations.  

In particular we look at (i) the impact of the Doppler and aberration effects on the CMB temperature and polarization power spectra and their corresponding cosmic variance, (ii) potential power and parity asymmetries induced in CMB maps as well as (iii) possible shifts in cosmological parameters in all-sky experiments. We also  investigate (iv) prospects of future experiments for measuring our local velocity w.r.t. the CMB using motion-induced harmonic mode coupling. Some of these issues have been studied previously, but here we redo them using state-of-the-art theoretical modeling of the boost and include polarization and cross-spectra which are usually ignored. 

We also introduce a software that offers an accurate boosting formalism called \texttt{CosmoBoost}\footnote{\href{https://github.com/syasini/CosmoBoost}{\faGithub ~ \texttt{syasini/CosmoBoost}}} which employs the generalized Doppler and aberration kernel developed in \citet{Yasini:2017jqg} based on previous work by \citet{Dai2014}. The calculations of the Doppler and aberration effect in this formalism are performed in harmonic space and hence the results are not prone to errors that incur in real space boosting due to pixelization and finite window function  \citep{Yoho:2012am,Jeong:2013sxy}. \texttt{CosmoBoost}\cite{cosmoboost_2019} also calculates the motion-induced spectral deviations of the CMB from black body in harmonic space \citep{Yasini:2017jqg}, which might also have an impact in deriving parameters from observations. However, these effects are expected to be small and would primarily impact other aspects of CMB data analysis (e.g. calibration) not considered here, and we therefore ignore them in this paper.

\textbf{Boost Variance:} The motion of the observer changes the uncertainty associated with determining the underlying theoretical power spectrum from observation. We will show that in a naive analysis of the power spectrum the boost changes the so called \emph{cosmic variance}; we call the resulting additional term  \emph{boost variance} and investigate its angular dependence and its relevance in different locations of the sky (see \S \ref{sec:boost_variance}).

\textbf{Power Asymmetry:} We will closely examine the effect of the Lorentz boost as a source of power asymmetry in the CMB \citep{Das:2018hnr,Shaikh:2019dvb,Dai:2013kfa, Aluri:2015tja, Aluri:2015cda} when observed in different portions of the sky, and in particular opposite hemispheres. This problem has been studied in \citet{Notari:2013} and \citet{Quartin:2014power} for CMB temperature; here we repeat the exercise using \texttt{CosmoBoost} and include polarization spectra as well. Using simulations we show that Lorentz boost induces non-trivial (percent level) hemispherical power asymmetry in TT, EE, and TE power spectra of a Planck-like experiment, and then compare the results with the observations from Planck \citep{Planck2018:isotropy} (see \S \ref{sec:hemispherical_power_asymmetry}).

\textbf{Parity Asymmetry:} Aside from the well-known power asymmetry, the Planck maps deviate from the expected statistical isotropy through the so called parity asymmetry, i.e., the odd multipoles carry more power than the even multipoles \citep{Aluri:2011wv}. Using simulations we  examine whether this could be caused by the motion of the observer \citep{Naselsky:2011jp} and investigate any potential parity asymmetry induced by the boost (see \S \ref{sec:parity_asymmetry}).

\textbf{Cosmological Parameter Estimation:} The motion-induced change in the power spectrum propagates to the inferred cosmological parameters as well. This issue has been reviewed in particular for the Planck parameter estimation in \citet{Catena:2012params} using only the temperature power spectrum, a simplified treatment of the foreground mask, and a suboptimal boosting scheme. Here we revisit the problem but also include (i) polarization, (ii) the final Planck noise configuration and foreground mask, and (iii) use the accurate boosting formalism \texttt{CosmoBoost} in simulating the power spectra (see \S \ref{sec:param_est}). Furthermore, the effect on high $\ell (>800$) and low $\ell (<800$) modes are examined separately to investigate potential discrepancies induced in these two ranges for Planck. The outcome of the combination of various updates with respect to \citet{Catena:2012params} is not immediately obvious---as some would lead to larger and others to smaller effects in parameter estimation. Therefore, it is worth redoing this exercise to ensure the Planck parameters are not biased by the Doppler and aberration effects.

\textbf{Boost Detection:} It is possible to infer the direction and amplitude of our local bulk motion with respect to the CMB by measuring the coupling between the nearby multipoles. This effect was introduced in \citet{Kosowsky2010}, and \citet{Amendola2010} based on the original calculations by \citet{Challinor2002}, and later detected by Planck in \citet{Aghanim:2013suk} using the CMB temperature. 
Since the amplitude of the coupling signal depends on the slope of fluctuation in the CMB angular power spectrum (see \S \ref{sec:boost_detection}), it actually leaves a larger signature in polarization than in temperature\footnote{The polarization power spectrum EE fluctuates with a larger slope than the temperature power spectrum TT.}. However, the polarization noise in Planck makes this component of the boost coupling signal subdominant with respect to the one in temperature, and hence undetectable. The next generation of CMB surveys, however, will have lower noise levels, and will therefore measure the polarization power spectrum better, and extend the range of modes measurable in temperature. Although the upcoming experiments such as the Simons Observatory (SO) \citep{SO:2020WP} and CMB-S4 \citep{CMB-S4:2020WP, CMB-S4:science2019} are planned to observe a smaller fraction of the sky compared with Planck, their higher sensitivities will map the power spectrum with greater precision at smaller scales, allowing for better measurement of the motion-induced mode coupling. It is therefore worthwhile to exploit their enhanced capabilities to make a more accurate measurement of our local velocity, independent of the dipole component. Such measurements will be extremely valuable in lifting the degeneracy between the kinematic and intrinsic components of the CMB dipole \citep{Yasini:2016dnd, Roldan:2016ayx}. We will examine the prospects of this detection with SO and CMB-S4, and the synergy between these experiments and Planck.\\

\textbf{Outline:} The outline of the paper is as follows: \S\ref{sec:notation} contains a summary of the notation and approximations used in the paper. In \S\ref{sec:boost_on_CMB} we review the Doppler and aberration Kernel formalism and then investigate the effect of Lorentz boost on the CMB temperature and polarization harmonic multipoles and power spectra as well as their associated cosmic variance. In \S\ref{sec:hemispherical_power_asymmetry} we examine the hemispherical power asymmetry induced by the boost in both temperature and polarization spectra. \S\ref{sec:parity_asymmetry} briefly inspects possible motion-induced parity asymmetries induced in CMB temperature and polarization power spectra. In \S\ref{sec:param_est} we examine the potential bias induced by the boost in cosmological parameter estimation for a Planck-like experiment. And finally in \S\ref{sec:boost_detection} the prospects of boost detection using large-sky CMB experiments such as SO and CMB-S4 are discussed. A summary of the results is presented in \S\ref{sec:summary}.


\section{Notation and approximations}\label{sec:notation} 
Throughout this paper, tilde $(~\tilde{~}~)$ and prime $(')$ accents are reserved for variables and indices in the boosted frame. The explicit use of the prime notation repeatedly throughout the text is to remind the reader that the variables of analysis are evaluated in a boosted frame. Whenever necessary the subscripts $r$ and $b$ are used to distinguish the \emph{rest} and \emph{boosted} frames w.r.t. the CMB. Unless noted otherwise,  $\Delta C_{\ell'} = \tilde{C}_{\ell'}-C_{\ell'}$ refers to the boosted minus rest frame values of the estimated power spectrum. A primed index on a rest frame observable such as $C_{\ell'}$ means $C_{\ell}|_{\ell=\ell'}$. $C^{XY}_\ell$ refers to the estimated power spectrum $1/(2\ell+1)\sum a^{X*}_{\ell m}a^{Y}_{\ell m}$, where $X,Y \in \{T,E\}$, and  $\C^{XX}_{\ell'}$ is used to indicate the underlying variance of the harmonic multipoles $\langle |a_{\ell,m}|^2 \rangle$. The estimated power spectrum is typically expressed with a hat notation ($\hat{C}_\ell$), however, we chose this unconventional notation (simply $C_\ell$) for the estimator to avoid the use of two accents (tilde and hat) for the estimated power spectrum in the boosted frame. Whenever the word \emph{cosmic variance} is accompanied by the symbol $\sigma$, it refers to the square root of $\langle (C_\ell - \langle C_\ell \rangle)^2 \rangle$.  

The fiducial values for the cosmological parameters used in the paper $(\omega_b, \omega_c, \theta, A_s, n_s,H_0,\sigma_8)$ are respectively $(0.022, 0.119, 1.042, 2.3\times10^{-9}, 0.9667, 67.74, 0.860)$. We also fix reionization $\tau=0.066$ to avoid complications in dealing with small $\ell$ values in parameter estimation. The values of cosmological parameters used are not the ones preferred by the most updated Planck 2018 results \citep{Planck:2018parameters}, however, since here we are only concerned with the amount of shift in these parameters induced by the boost, their actual values are of little importance in this study. Finally, $h$, $k$, and $c$ respectively refer to the Planck, Boltzmann, and speed of light in vacuum constants. 

We use the Planck 143 GHz mask\footnote{\href{http://pla.esac.esa.int/pla/\#maps}{http://pla.esac.esa.int/pla/\#maps}} as the primary mask in the study. The 100 GHz and 217 GHz masks have respectively smaller and larger areas covered with respect to the 143 GHz. This means that we should expect the boost effects in the power spectrum to be slightly weaker for the former and stronger for the latter \citep{Pereira2010}. For this reason, we chose to use the 143 GHz mask to get a reasonable estimate of the effect in the combined maps. 

In the calculations of the aberration kernel and its analytical estimates we do not account for the frequency dependence of the boost because it is negligible for the cases studied here (especially for experiments with relatively symmetric masks such as Planck). These effects can become non-trivial for partial-sky surveys though and should be taken into account. Additionally, we neglect any effects due to extragalactic foregrounds \citep{Chluba:2004vz, Balashev:2015lla}. Also, 
for simplicity we do not consider the CMB B-mode power spectrum throughout the paper (see \citet{Yasini:2017jqg} for details on motion-induced E to B leakage).

\section{The effect of our motion on temperature and polarization statistics}\label{sec:boost_on_CMB}

\subsection{Doppler and Aberration Kernel Formalism}

\begin{figure}
    \centering
    \includegraphics[width=\linewidth]{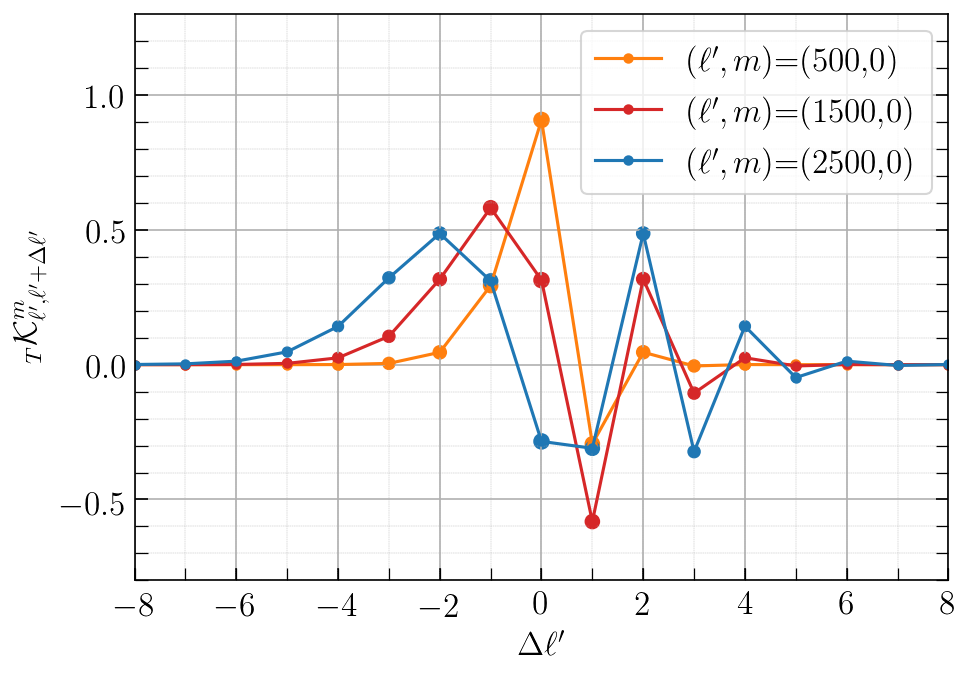}
    \caption{The general behavior of the Doppler and aberration kernel for thermodynamic temperature $T$. The kernel coefficient $\Ker{T}{\ell'}{\ell}$ denotes the amount of motion-induced leakage from the spherical harmonic coefficient $a_{\ell m}$ in the rest frame, into the observed $a_{\ell' m}$.  The $y$-axis shows the leakage components from a range of $-8 < \Delta \ell' < 8$ neighbors into observed modes $\ell'=$500, 1500, and 2500 at $m=0$. As we increase $\ell'$, the kernel becomes more non-linear and the contribution from farther neighbors become more important and hence non-negligible. The kernel coefficients generally become smaller as $m \rightarrow \ell'$. Also, at $\ell' \gg 2$ the polarization kernel $\Ker{E}{\ell'}{\ell}$ converges to $\Ker{T}{\ell'}{\ell}$. }
    \label{fig:Kernel_shape}
\end{figure}

The effects of the boost on the incoming photons in the observer's frame moving with the peculiar velocity vector $\vec{\textbf{\textrm{v}}}$ w.r.t. the CMB are two fold: First, there is a change in the frequency of the photons $\nu$ due to the Doppler effect 

\begin{equation}\label{eq:doppler}
\nu'=\gamma (1+\beta \mu)\nu,
\end{equation}
and second, there is a change in the observed angle of the photons $\gammahat$ due to the aberration effect  
\begin{equation}\label{aberration}
\gammahat'=\Big(\frac{(1-\gamma^{-1})\mu+\beta}{1+\beta \mu}\Big )\betahat+\Big (\frac{\gamma^{-1}}{1+\beta \mu}\Big)\gammahat.
\end{equation}
Here $\gamma=1/\sqrt{1-\beta^2}$, $\mu=\gammahat \cdot \betahat$, and $\vec{\bm\beta}=\beta \betahat = \vec{\textbf{\textrm{v}}}/c$ is the dimensionless velocity of the frame. For simplicity, it is conventional to work in the coordinate system where  $\hat{\bm z}=\betahat$ and rewrite Eq.~\eqref{aberration} as 
\begin{equation}\label{eq:aberration_z}
\mu'=\frac{\mu+\beta}{1+\beta \mu}
\end{equation}
where $\mu'=\gammahat' \cdot \betahat$. Henceforth, we use this coordinate system for all the analyses and rotate all sky maps to this frame whenever necessary. Using the transformation $\beta \rightarrow -\beta$, one easily obtains the inverse transformations $\nu \rightarrow \nu'$ and $\mu \rightarrow \mu'$ to find the variables in the CMB frame. 

Assuming the CMB is a perfect black body in its rest frame in every direction with the specific intensity 

\begin{equation}
I_\nu(\gammahat) = \frac{2 h }{c^2} \frac{\nu^3}{e^{h \nu/k T(\gammahat)}-1},
\end{equation}
it is easy to show, using the inverse of Eq.~\eqref{eq:doppler}, that the temperature in the observer's moving frame is equal to

\begin{equation}
    \tilde{T}(\gammahat') = \frac{T(\gammahat)}{\gamma(1-\beta\mu')},
\end{equation}
and similarly for the polarization Stokes parameters $Q_T$ and $U_T$ in thermodynamic temperature units. Decomposing this equation in spherical harmonic space yields the harmonic boost equation \citep{Yasini:2017jqg, Dai2014}:

 \begin{equation}\label{eq:alm_kernel}
 \tilde{a}^X_{\ell' m} =
 \sum_{\ell}~_X\mathcal{K}^{m}_{\ell' \ell}(\beta) ~a^X_{\ell m}.
 \end{equation}
Here $X \in \{T, E \}$ and $_X\mathcal{K}^{m}_{\ell' \ell}(\beta)$ is the Doppler and aberration kernel (see \S \ref{sec:app:kernel} for full definition) . The kernel coefficient $\Ker{X}{\ell'}{\ell}$ represents the amount of motion-induced leakage of multipole $a_{\ell m}$ from the rest frame, into the observed $\tilde{a}_{\ell' m}$ in the moving frame. Obviously, when $\beta = 0$, the kernel reduces to the Kronecker delta $\Ker{X}{\ell'}{\ell}(0)=\delta_{\ell' \ell}$. Note that there is no cross-leakage among different $m$ modes. This simplification arises because of the $\betahat = \hat{\bm z}$ assumption, since in this frame the angular distortions due to the aberration effect are only in the direction of the polar angle $\hat{\bm \theta}$, with no change in the azimuthal direction $\hat{\bm \phi}$. 

Fig. \ref{fig:Kernel_shape} shows the shape of the kernel $\Ker{T}{\ell'}{\ell'+\Delta \ell'}$ for a few different  multipoles $\ell'$, and a range of $-8 < \Delta \ell' < 8$. Each point represents the amount of motion-induced leakage from multipole $\ell'+\Delta \ell'$ into the observed $\ell'$ mode. In the moving frame, a multipole $\ell'$ observed in the hemisphere towards the direction of motion leaks into its higher neighbors ($\ell'+1, \ell'+2,$ etc.) and receives a contribution from its lower neighbors ($\ell'-1, \ell'-2,$ etc.). And the opposite happens in the other hemisphere. With this in mind, one can interpret the $\Delta \ell'<0$ ($\Delta \ell'>0$) terms in the kernel as the contribution of modes towards (away from) the direction of motion.

\begin{figure*}
\centering
\centering
    \includegraphics[width=0.55\linewidth]{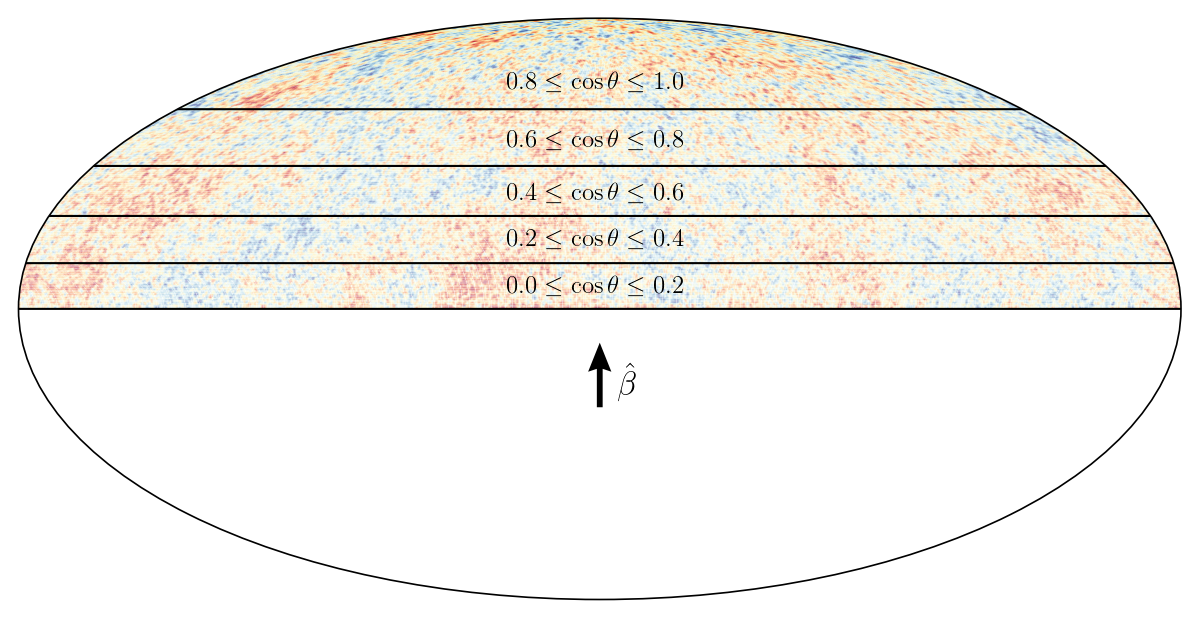}
\begin{tabular}{c  c }
    \includegraphics[width=.49\textwidth]{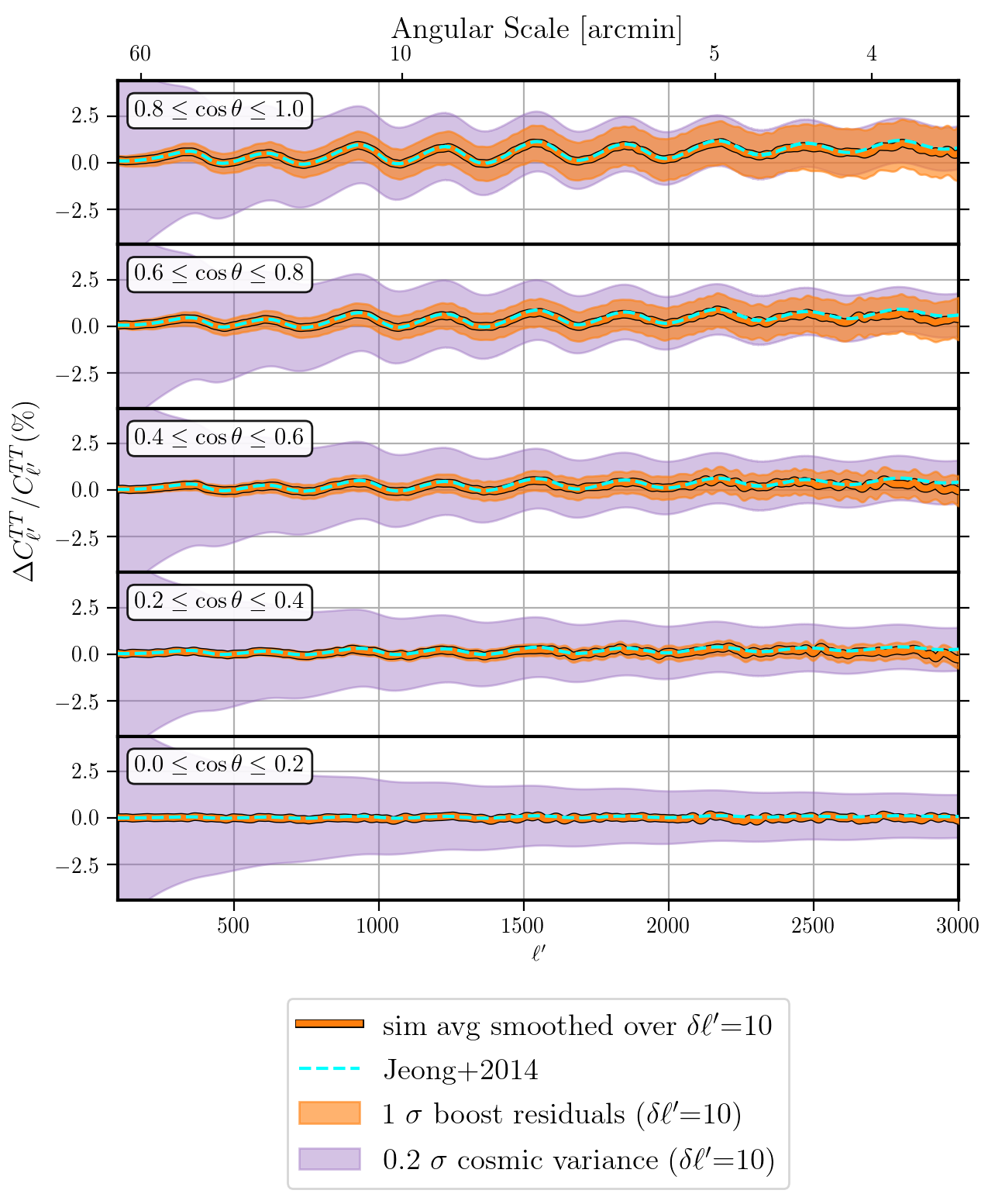} &
    \includegraphics[width=.49\textwidth]{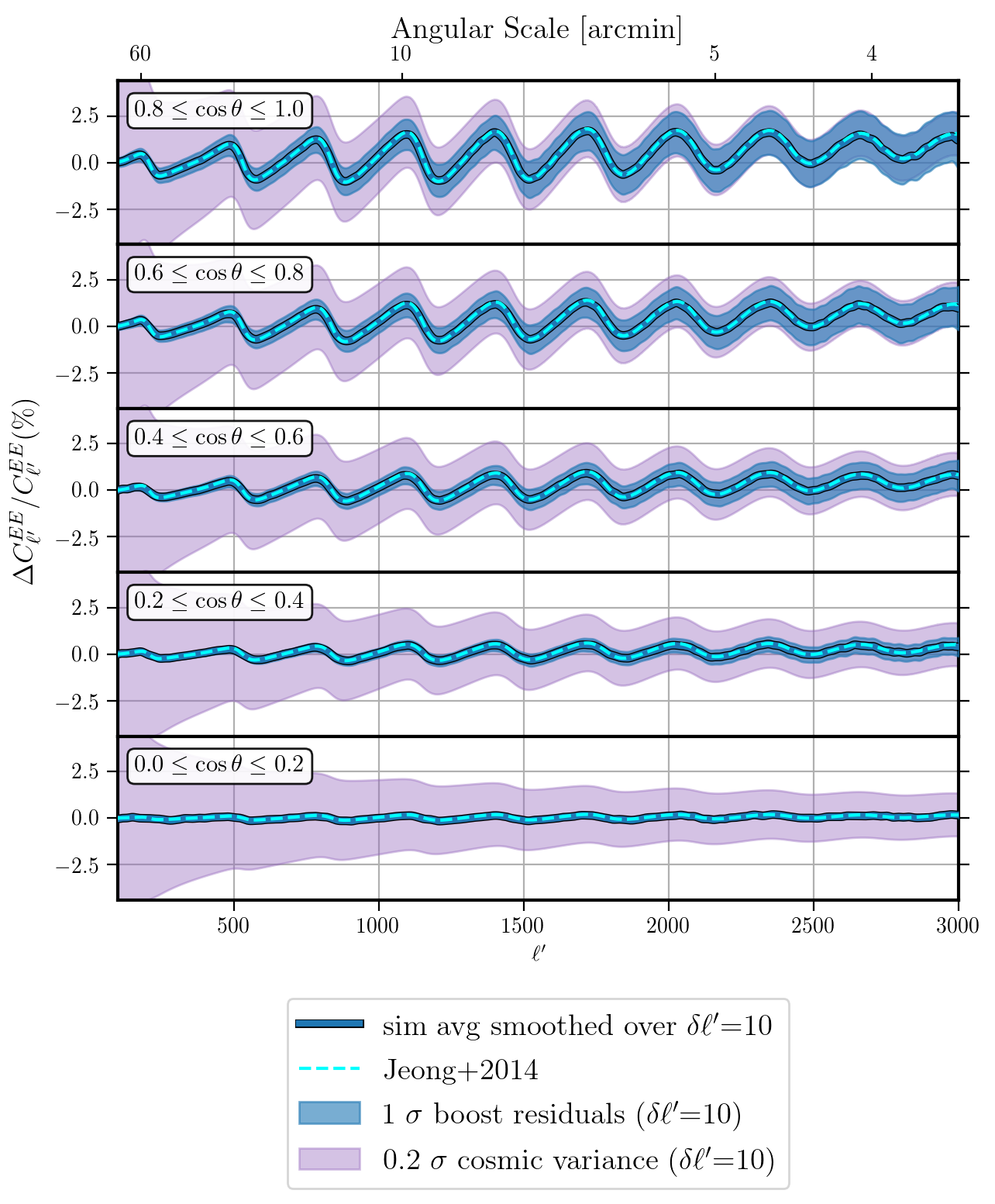} \\
    \small (a) Temperature & \small (b) Polarization
  \end{tabular}
\caption{Induced relative change in the CMB power spectra ($\Delta C_{\ell'}/C_{\ell'} = \tilde{C}_{\ell'}/ C_{\ell'} -1$) due do the Doppler and aberration effects induced in a frame moving towards the northern galactic pole. Individual plots respectively show this change for TT \emph{(left)} and EE \emph{(right)}, in 5 equal-area strips in the northern hemisphere \emph{(top)}. The thick solid lines (\emph{orange} for TT and \emph{blue} for EE) show the average of 100 simulations Gaussian smoothed over a $\delta \ell'=10$ scale. The effect becomes larger towards the direction of motion (highest panels in both plots) and in general is more prominent in EE than in TT. The analytical formula from Jeong+14 \emph{(dashed cyan)} emulates the average effect extremely well, but if used for correcting individual power spectra it leaves residuals in the data. The \emph{shaded bands} (\emph{orange} for TT and \emph{blue} for EE) around the simulation average, show the 1$\sigma$ region for the boost residuals left in individual realizations of the power spectrum, which can be as large as 20\% of cosmic variance (\emph{purple band}) at $\ell \simeq 3000$ for both TT and EE (see Eq.~\ref{boost_residual}). }
\label{fig:5strip}
\end{figure*}

At low $\ell'$ the kernel starts very linearly ($\approx \beta \ell'$) with a sharp peak at $\Delta \ell'=0$ \citep{Challinor2002}. But as we pass $\ell'\gtrsim 1/\beta \approx 800$, the behaviour of the kernel becomes non-linear (deviates from $\beta \ell' $ approximation) and the central element $\Delta \ell'=0$ does not have the highest value anymore \citep{Chluba2011}. In other words, at this range the leakage from farther neighbors becomes far greater than one would naively expect. For the high values of $\ell'$ presented in the plot the kernel for polarization is practically the same; the difference between the temperature and polarization kernel coefficients is of the order $\sim 1/(\ell^2-2)$.

\subsection{Boosted Power Spectrum} \label{sec:boosted_PS}
\subsubsection{Analytical estimate}
The harmonic boost equation (Eq.~\eqref{eq:alm_kernel}) indicates that in the observer's frame the multipole $\ell'$ comprises the mode $\ell'$ of the rest frame, as well as the motion induced leakage from the first neighbors $\ell'+1$ and $\ell'-1$, second neighbors $\ell'+2$ and $\ell'-2$, and so on. The leakage from the nearby modes leads to a mode coupling between multipoles as well as a change in the estimated power spectrum 

\begin{equation}
    \tilde{C}^{XX}_{\ell'} = \frac{1}{2\ell'+1}\sum_{\ell,m}\left| \Ker{X}{\ell'}{\ell}(\beta) \right|^2 C^{XX}_\ell.
\end{equation}
\citet{Jeong:2013sxy} has shown that the ensemble average of this expression can be well approximated by

\begin{equation}\label{eq:Jeong1}
    \tilde \C^{XX}_{\ell'} = \C^{XX}_{\ell'} - \beta \ell' \overline{\cos\theta'} \frac{\diff \C^{XX}_{\ell'}}{\diff \ell'} + \mathcal{O}(\beta^2).  
\end{equation}
where $\C^{XX}_{\ell'}  = \langle C^{XX}_{\ell'} \rangle $ and $\theta' = \arccos \mu'$ is the polar angle in the moving frame 
and the overline indicates an angular average

\begin{equation}
    \overline{\cos\theta'} = \frac{\int_{\Omega_{\rm obs}} \cos \theta' \sin \theta' \diff \theta' \diff \phi'}{\int_{\Omega_{\rm obs}} \sin \theta' \diff \theta' \diff \phi'},
\end{equation}
where $\Omega_{\rm obs}$ is the observation patch in the sky. In the following subsection we will show that Eq.\ref{eq:Jeong1} well approximates  
the statistical average of the motion-induced shift in the power spectrum, but it does not properly correct all effects of the boost on an individual realization.

\subsubsection{Simulations}

Using the formalism presented in the previous subsection, we now examine the amount of motion-induced distortion in the CMB power spectra using 100 rest frame and boosted simulations. In order to gauge the significance of the effect in the observed CMB sky, we follow the example in \citet{Jeong:2013sxy} and split the northern hemisphere into 5 equal-area strips (assuming the observer is moving towards $\hat{\bm z}$). Then we look at the relative difference between the rest frame and boosted frame power spectra in each of these strips for an individual simulation, as well as the average of the ensemble.

\begin{figure*}
\centering
\begin{tabular}{c  c }
    \includegraphics[width=.49\textwidth]{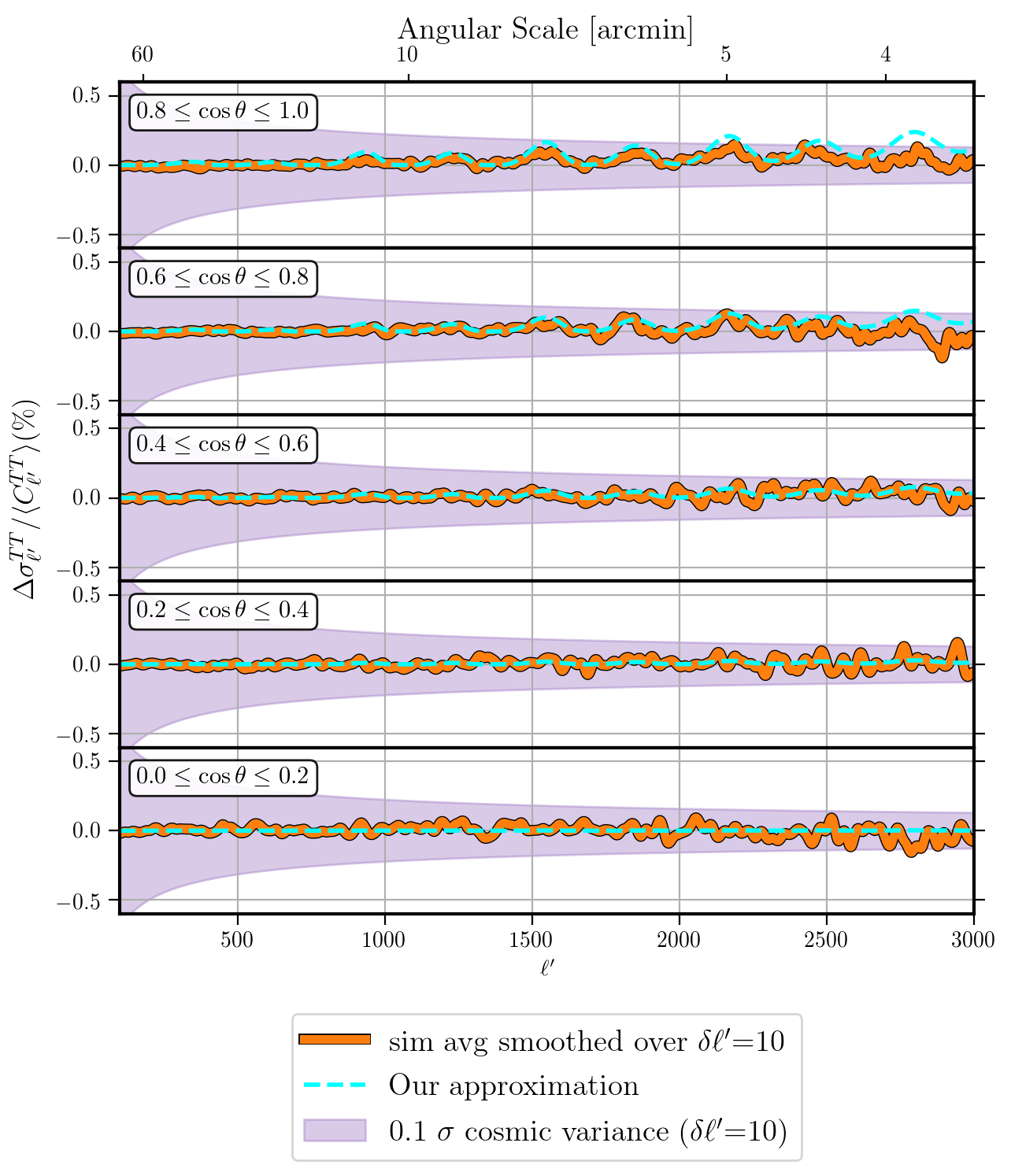} &
    \includegraphics[width=.49\textwidth]{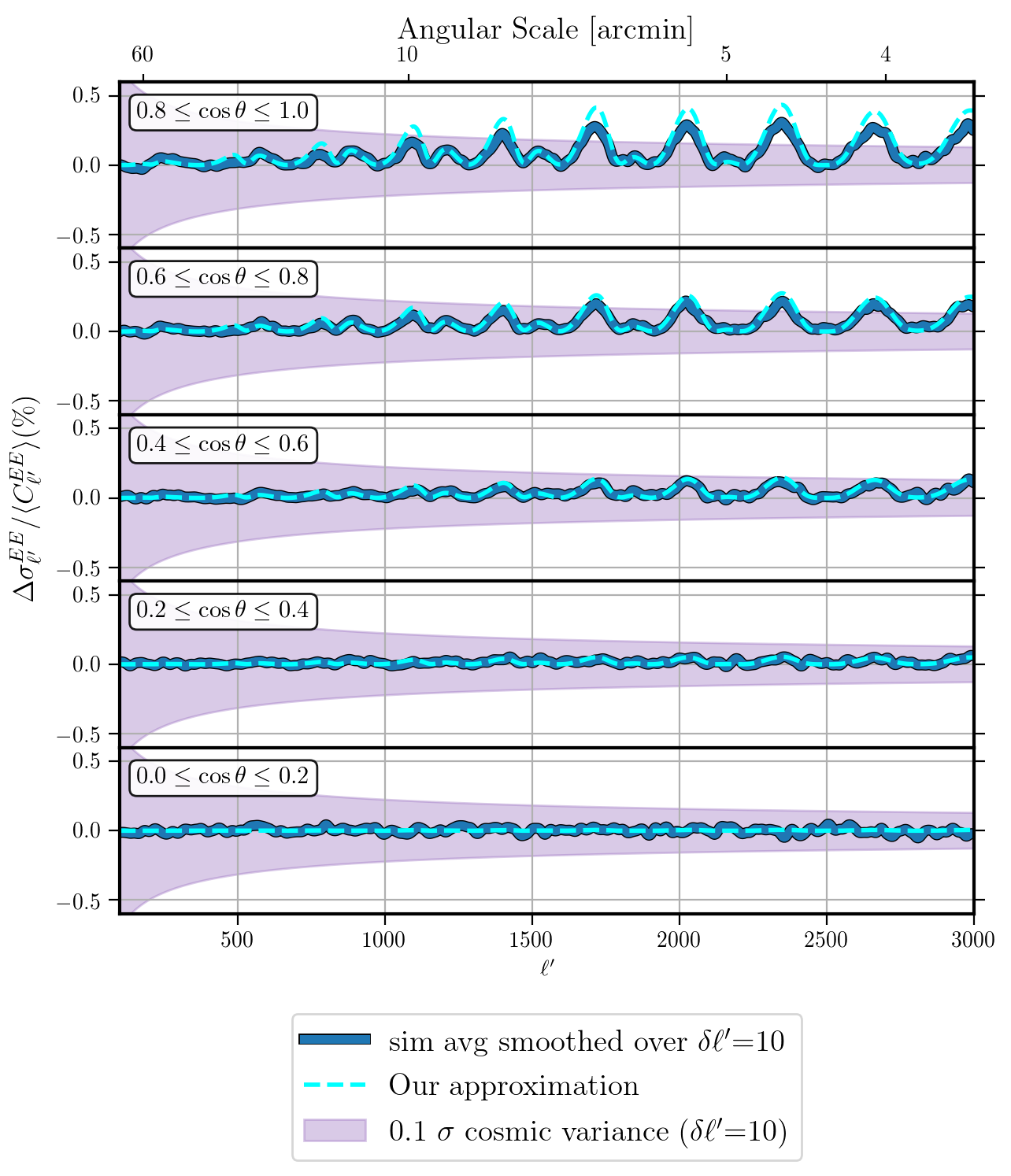} \\
    \small (a) Temperature & \small (b) Polarization
  \end{tabular}
\caption{\emph{Boost variance} (shift in the cosmic variance due to the Doppler and aberration effects) in the 5 equal area strips of Fig. \ref{fig:5strip}, normalized by the rest frame power spectrum. Individual plots respectively show this change for TT \emph{(left)} and EE \emph{(right)} spectra. The solid lines (\emph{orange} for TT and \emph{blue} for EE) show the average difference between cosmic variance in the boosted frame $\tilde{\sigma}_\ell$ (Eq.~\ref{eq:boost_variance}) and rest frame $\sigma_\ell$ (Eq.~\ref{eq:cosmic_variance_text}) for 100 simulations Gaussian smoothed over a $\delta \ell'=10$ scale. The increase due to the boost variance can be as large as 10 \% of rest frame cosmic variance (\emph{purple band}) for TT, and 20\% for EE at $\ell \simeq 3000$. Similar to what happens to the average of the power spectrum shown in the previous plot, the variance also becomes larger towards the direction of motion (higher panels in both plots) and in general is more prominent in EE than in TT. Unlike the average, however, the variance does not change sign in the southern hemisphere (not shown here). Our analytical formula in Eq.~\eqref{eq:boost_variance} approximates the effect closely, but slightly over estimates it at its peaks.}
\label{fig:5strip_BV}
\end{figure*}

Fig. \ref{fig:5strip} shows the motion-induced fractional change $\Delta C_{\ell'}/C_{\ell'}$ in 5 equal-area bands in the northern hemisphere for TT and EE power spectra. There are obvious general characteristics in the plot that are expected for the ensemble average of the boost according to Eq. \ref{eq:Jeong1}: (i) The effect is more prominent in the strips closer to the north galactic pole because the Doppler and aberration effects are stronger in directions closer to the apex of motion. (ii) In the northern hemisphere, there is an overall increase in the power that becomes larger as we get to smaller angular scales (higher $\ell$ modes)---this is a direct consequence of the fact that the aberration effect changes the relative size of smaller anisotropies more strongly than it does for large scales\footnote{To put it simply and perhaps more intuitively, it takes more aberration effect to make the quadrupole look like the octupole ($\ell=2 \rightarrow \ell'=3$) than it does for $\ell=1000 \rightarrow \ell'=1001$. Therefore, for the same $\beta$ the former would be a smaller effect than the latter---i.e., aberration affects smaller scales more strongly than large scales}. (iii) Since the amount of power leakage depends on the slope of the power spectrum, the relative changes in TT and EE fluctuate out of phase with respect to the corresponding rest frame spectra (second term in Eq.  \ref{eq:Jeong1}). By the same token, since the EE has more pronounced fluctuations than the TT power spectrum, the relative change is in general larger in polarization than in temperature. As we will discuss in more detail in \S \ref{sec:boost_detection}, for an experiment with low instrumental noise, this very fact makes detection of the boost---using mode coupling---easier in polarization than in temperature (see Appendix \ref{sec:app:TE} for the TE power spectrum plots). It is also worth noting that although the behavior of the kernel becomes nonlinear in $\beta$ for $\ell\gtrsim1/\beta \simeq800$, Eq.~\eqref{eq:Jeong1} still performs very well in approximating the ensemble average of the boosted power spectra in this range. 

Another feature that can be seen in Fig.~\ref{fig:5strip} is that the power in an individual mode $\ell'$ of the boosted realization ($\tilde C_{\ell'}$) may differ substantially from the rest frame one ($C_\ell$) beyond what is represented by the ensemble average in Eq.~\ref{eq:Jeong1}. Statistically the effect of the boost in the map $\Delta C_{\ell'}$ deviates from the ensemble average $\Delta \C_{\ell'}$ by\footnote{We  assume that Eq. \ref{eq:Jeong1} holds for an individual realization $\tilde{C}_{\ell'}$.} 

\begin{align}\label{boost_residual}
    \langle (\Delta C_{\ell'} - \Delta \C_{\ell'})^2 \rangle =&
    (\beta \ell \overline{\cos \theta})^2 \langle (\frac{\diff C_{\ell'}}{\diff \ell'}  - \frac{\diff \C_{\ell'}}{\diff \ell'})^2 \rangle \nonumber\\
    \simeq&  \frac{(\beta \ell \overline{\cos \theta})^2}{2\ell'+1}\C^2_{\ell'},
\end{align}
 where the second equality is derived heuristically. Eq.~\ref{boost_residual} indicates that using the ensemble average to correct the motion-induced effects on $\tilde C_\ell$ would leave a non-negligible amount of residuals in the power spectrum. As depicted in the top panels of both plots in Fig. \ref{fig:5strip}, the residuals left in a single realization of the power spectra binned by $\delta \ell' =10$ can be as large as 20\% of the rest frame cosmic variance ($\sigma$) on a wide range of angular scales. These residuals can be potentially important for small-area surveys such as ACTPol which uses the Jeong approximation to correct for the Doppler and aberration effects (see \S \ref{sec:practical_guide}). As pointed out earlier, by using the accurate formalism of \texttt{CosmoBoost} which corrects the motion-induced effects at map-level in harmonic space, these residuals can be entirely removed. Another strategy for removing these residuals---similar to what is done for cosmic variance---would be to employ a large bin size when correcting the effect statistically; they virtually vanish for $\delta \ell'>50$, however, their ratio w.r.t. cosmic variance remains unchanged.

An important point that might not be immediately obvious from Fig. \ref{fig:5strip} is that the oscillating patterns emerge in the southern hemisphere with the opposite sign ($\cos\theta \rightarrow -\cos\theta : \Delta C_\ell/C_\ell \rightarrow -\Delta C_\ell/C_\ell$ in $\mathcal{O}(\beta$) ). The crucial consequence of this statement is that when one calculates the power spectrum over a patch of the sky that is symmetric w.r.t. $\cos\theta$, the motion-induced effects cancel each other to first order in $\beta$. 
As we will show in \S \ref{sec:param_est}, this is roughly the case for Planck which has a mask that is fairly symmetric w.r.t. the peculiar velocity unit vector $\betahat$ (also see \S IV.B in \citet{Jeong:2013sxy}).

\subsection{Boost Variance}\label{sec:boost_variance}
\subsubsection{Analytical estimate}
Aside from inducing a shift in the statistical average of the power spectrum at every $\ell$, the Doppler and aberration effects also change the variance of each mode. In what follows we will quantify this variance analytically. For simplicity we drop the $XX$ superscript from all $C^{XX}_\ell$s and $\C^{XX}_\ell$s, and the prime (~\'~) accent in this subsection.

We are interested in measuring the variance of the (CMB rest frame) harmonic coefficients $\langle |a_{\ell m}|^2 \rangle\equiv\C_\ell$, also known as the theoretical power spectrum of the underlying cosmological model.
The power spectrum estimated from the map \citep{Tegmark:1997vs} $C_\ell = \sum_m |a_{\ell m}|^2 /(2\ell+1)$ has the following first cumulant (ensemble mean):

\begin{equation}
    \langle C_\ell \rangle = \C_\ell.
\end{equation}
Similarly, for the second cumulant we have (see appendix \ref{sec:app:cosmic_variance} or \ref{sec:app:moments_of_Cl})
\begin{equation}\label{eq:cosmic_variance_text}
    \sigma^2_\ell = \langle (C_\ell-\C_\ell)^2 \rangle = \frac{2}{2\ell+1}\C^2_\ell
\end{equation}
which is known as \emph{cosmic variance} \citep{Knox:1995dq, Scott:2016fad}. A simple interpretation of this equation is that in every ensemble of power spectrum realizations with the mean value $\C_\ell$, at every $\ell$ roughly 68\% 
of the $C_\ell$s are within $\sigma_\ell$ of the mean.

In a moving frame, an observed multipole $\ell'$ leaks into and receives a contribution from its nearby neighbors ($\ell\pm1, \ell\pm2,$ etc.). This phenomenon, which roughly depends on the slope of the power spectrum (see Eq.~\eqref{eq:Jeong1}), increases the uncertainty associated with cosmic variance: not only the variance of each mode depends on the mean $\C_\ell$, it also weakly depends on the slope $\diff \C_\ell/\diff \ell$ because of the observer's motion. We call this source of extra uncertainty induced by the observer's motion \emph{boost variance}. A simple calculation of the squared deviation of the boosted power spectrum $\tilde{C}_\ell$ from the theoretical $\C_\ell$ yields (see appendix \ref{sec:app:boost_variance})

\newsavebox\itemA
\savebox\itemA{$\tilde{\sigma}^2_\ell = \langle (\tilde{C}_\ell-\C_\ell)^2 \rangle $}

\newsavebox\itemC
\savebox\itemC{$\simeq \dfrac{2}{2\ell+1} \C^2_\ell$}

\newsavebox\itemB
\savebox\itemB{$\left(\beta\ell \overline{\cos\theta} \dfrac{\diff \C_\ell}{\diff \ell}\right)^2$.}

\[
\def\stackalignment{l}
\stackon{%
  \boxed{ \usebox{\itemA}  \usebox{\itemC}+ \usebox{\itemB} }
}{%
  \phantom{\kern\wd\itemA}^{} \phantom{{}={}}
  \overbrace{\kern\wd\itemC}^{\text{cosmic variance}}
  \overbrace{\kern\wd\itemB}^{\text{boost variance}}
}%
\label{eq:boost_variance} 
\tag{13}\]
The motion of the observer increases the variance of the ensemble of observed power spectra both towards ($\overline{\cos\theta}>0$) and away from ($\overline{\cos\theta}<0$) the direction of motion, and has minimal effect on the motion's equatorial plane ($\overline{\cos\theta}\simeq 0$). In a partial sky experiment, the sample variance can be easily calculated from this expression via $\tilde \sigma^2_\ell /f_{\rm sky}$  where $f_{\rm sky}$ is the fraction of sky covered by the survey \citep{Scott:1993yv}. 

\subsubsection{Simulations}\label{subsubsec:boost_var_simulations}
Fig. \ref{fig:5strip_BV} shows the significance of boost variance for the 5 equal area cuts of Fig. \ref{fig:5strip}. Here the \emph{left} and \emph{right} plots show the relative motion-induced change in the standard deviation $\Delta \sigma^{XX}_\ell/\langle C^{XX}_\ell \rangle = (\tilde{\sigma}^{XX}_\ell - \sigma^{XX}_\ell)/\langle C^{XX}_\ell\rangle$ respectively for temperature ($X=$ T) and polarization ($X=$ E). As expected, the change due to boost variance is larger towards the direction of motion (higher panels) and smaller angular scales (higher $\ell)$, and the effect is generally larger in polarization than in temperature. In the topmost panel which represents $f_{\rm{sky}}=10\%$ of the sky towards $\betahat$, the effect is around 10\% of rest frame cosmic variance ($\sigma_\ell$) in TT and 20\% in EE.

\begin{figure}
    \centering
    \includegraphics[width=0.99\linewidth]{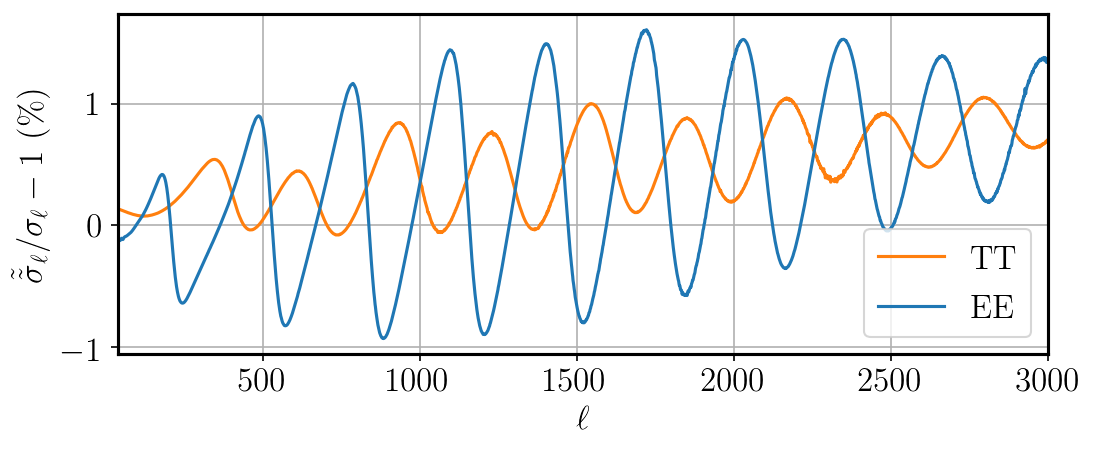}
    \caption{Relative change in cosmic variance due to the boost after correcting the mean power spectrum $\C_\ell \rightarrow \tilde{\C}_\ell$. }
    \label{fig:BV_corrected}
\end{figure}

It is important to point out that Eq.~\eqref{eq:boost_variance} represents the relevant variance in a naive analysis of the power spectrum without any boost corrections implemented (e.g., in a study where the theoretical power spectrum or cosmological parameters are being inferred from the observed $\tilde{C}_\ell$ in a moving frame, without correcting the Doppler and aberration effects). However, if the effect of the boost is already being corrected for in the theoretical mean $ \C_\ell \rightarrow \tilde{\C}_{\ell}$, as is done in e.g. SPTPol (see \S 7.4 in \cite{Henning:2017nuy}) then the variance reduces to
\stepcounter{equation}
\begin{equation}\label{eq:boost_variance_corrected}
    \tilde{\tilde{\sigma}}^2_\ell=\langle (\tilde{C}_\ell - \tilde{\C}_\ell)^2\rangle \simeq \frac{2}{2\ell+1}\tilde{\C}^2_{\ell} = \frac{2}{2\ell+1}(\C_{\ell} - \beta \ell \overline{\cos\theta} \frac{\diff \C_{\ell}}{\diff \ell}  )^2.
\end{equation}
After correcting the effect statistically, this would be the proper expression to use for cosmic variance in the boosted frame. Expanding the above equation in $\beta$ and subtracting the rest frame cosmic variance $2\C^2_{\ell}/(2\ell+1)$ yields a boost variance equal to $\simeq -4\beta \overline{\cos \theta} \C_\ell \diff \C_\ell /\diff \ell$. Here the ``boost variance around the boosted mean'' is about an order of magnitude smaller than the expression in Eq.~\eqref{eq:boost_variance} which shows the boost variance around the unboosted mean. 
Fig. \ref{fig:BV_corrected} shows the relative difference of this extra variance w.r.t. cosmic variance ($\sigma$).

The motion-induced effects in the power spectrum change the mean and variance of the inferred cosmological parameters. Here we estimate the change in the variance using Fisher matrix.
By propagating the variance in Eq. \eqref{eq:boost_variance_corrected} to inferred cosmological parameters via the Fisher matrix ${{\bm F}}_{ij}=\sum_\ell \tilde{\C}_\ell^{-2} \partial_i\tilde{\C}_\ell\partial_j\tilde{\C}_\ell$ \citep{Tegmark:1997vs} up to $\ell_{\rm max} = 5000$, we found that the error bars on all parameters are affected at most by $0.015 \sigma$ in both TT and EE, assuming $\overline{\cos\theta}=1$; this only includes the change in variance and not the shift in parameters which can be an order of magnitude larger. Therefore correcting the boost in the average (or theoretical) power spectrum should sufficiently remove any bias propagated to the inferred  parameters from $\tilde{C}_\ell$.  


\subsection{Accounting for the Boost: Practical Guide}\label{sec:practical_guide}

After the theoretical insights of the previous sections, it is sensible to ask what is the best strategy to correct for the boost in using the CMB to derive  cosmological parameters.
Here is a summary of possible  deboosting methods:

\begin{enumerate}[label=\arabic*)]

    \item \textbf{Real space deboosting}: The Doppler and aberration effects can be corrected at map level on individual pixels. This approach  potentially underestimates the boost effect at small angular scales, and is sensitive to the pixel window function \citep{Yoho:2012am}. 
    
    \item \textbf{Harmonic space deboosting}: The spherical harmonic coefficients $a_{\ell m}$ can be deboosted with the Doppler and aberration kernel formalism of \texttt{CosmoBoost}. This is the most accurate way to deboost the CMB without leaving any residuals in the average or variance of $C_\ell$ \citep{Yasini:2017jqg}.

    \item \textbf{Corrections to the power spectrum}: Eq. \ref{eq:Jeong1} \citep{Jeong:2013sxy} provides an excellent estimate for the effect of the boost on the theoretical power spectrum $\C_\ell$. If the analysis demands a comparison between the observed (boosted) power spectrum $\tilde{C}_{\ell'}$ and the theoretical $\C_\ell$---e.g., likelihood analysis in parameter estimation---there are two ways this equation can be used to correct for the boost:
    (i) boost the theoretical power spectrum for a given model $\C_\ell \rightarrow \tilde{\C}_\ell$ directly, and (ii) use the ensemble average $\tilde{\C}_{\ell'}$ to deboost the observed power spectrum $\tilde{C}_{\ell'}$. In both case the correct variance to use is represented by Eq.~\ref{eq:boost_variance_corrected}.\\

\textit{Note}: If no boost correction is performed on the observed $\tilde{C}_{\ell'}$ at all, the average and variance are biased respectively by Eqs. \ref{eq:Jeong1} and \ref{eq:boost_variance}.

\end{enumerate}

Assuming the amplitude and direction of the boost is known, the most accurate way to deboost an observed map is to use \texttt{CosmoBoost}. In this approach, the individual spherical harmonic coefficients are deboosted and no residuals are left neither in the map nor in the estimated power spectrum. Moreover, the final product can be reliably used in cross--correlation analysis. The downside is that calculating and storing the kernel coefficients can become numerically prohibitive at very high $\ell$. 

Alternatively, if one is only interested in the power spectrum $C_{\ell}$ (and not the map or $a_{\ell m}$s), one can use the Jeong+14 formula to correct for the boost effects. In such a case, it is advisable to apply Eq. \ref{eq:Jeong1} to the theoretical (rest frame) power spectrum for a given set of parameters before comparing to the data. The variance in this case is given by Eq.~\ref{eq:boost_variance_corrected}.

Which experiments should care to explicitly correct for the boost?
The CMB power spectrum of whole-sky and/or symmetric-sky experiments (e.g. Planck and SO) will not be impacted by the boost: the dependence of the observed spectrum  on $\overline{\cos\theta}$ ensures that the corrections are negligible ($\overline{\cos\theta}\simeq 0$).
However,  should a smaller area  of the experiment be used in cross correlation analysis with other types of surveys, it might be appropriate to deboost, given the effect of the boosting on individual $\ell$ modes (see Fig. \ref{fig:5strip}). 

In partial-sky experiments, the effects of the boost should be evaluated on the basis of  where the area observed is located on the sky, as well as  the attainable precision in cosmological parameter estimation (also influenced by instrumental characteristics like beam size and noise level).
If no boost correction is applied (e.g. BICEP, POLARBEAR)
the average and variance of the power spectrum are biased according to Eqs. \ref{eq:Jeong1} and \ref{eq:boost_variance}, which can be used to assess the approximation made.

Other experiments have used the Jeong+14 formula to try and correct for the Doppler and aberration effects.
ACTPol \citep{Louis:2016ahn} uses the boosted ensemble average $\tilde{\C}_\ell$ to subtract the average effect from the observed power spectrum $\tilde{C}_{\ell'}$.
SPTPol \citep{Henning:2017nuy} opted for the strategy of boosting the theoretical ${\C}_{\ell}$ before comparing it with the observational points in the likelihood analysis. In both these cases, the standard (rest frame) cosmic variance should be replaced with Eq. \ref{eq:boost_variance_corrected}. In general, if map level correction with \texttt{CosmoBoost} is not feasible, the latter approach is recommended since it introduces minimal error due to leftover boost effects (\S\ref{sec:boosted_PS}).

\section{hemispherical Power asymmetry in the power spectrum}\label{sec:hemispherical_power_asymmetry}

Since the Doppler effect increases the brightness of the incoming radiation on all angular scales in the direction of motion---and decreases it on the opposite side---it creates a power asymmetry in the two hemispheres. It was discovered in \citet{Eriksen:2003db} that the WMAP power spectrum calculated over different patches of the sky does not seem to be statistically isotropic. This anomaly was  later confirmed in the Planck maps as well \citep{Planck2013:isotropy,Planck2015:isotropy,Planck2018:isotropy}. It is worth looking into how the motion of the observer affects power distribution in the CMB, and possibly shed light on the hemispherical asymmetry anomaly observed in the CMB map. This exercise has been performed for the CMB temperature in \citet{Notari:2013} and \citet{Quartin:2014power}, but here we repeat it using the accurate formalism of \texttt{CosmoBoost}, and also include polarization (see \citet{Mukherjee:2013zbi} for an analytical examination using Bipolar Spherical Harmonics \citep{Hajian:2003qq}).    

As discussed in the previous section, the observed power spectrum increases in the direction of motion proportional to $\beta \ell' \overline{\cos \theta'} \diff C_{\ell'}/\diff \ell'$ (see Fig. \ref{fig:5strip}). Since $\overline{\cos \theta'}$ is positive in the northern hemisphere and negative in the southern hemisphere, this naturally lead to a power asymmetry in the observed CMB anisotropies both in temperature and polarization. In order to gauge the amount of motion-induced asymmetry in the CMB sky we take the following steps: we split the Planck foreground mask into northern and southern hemispheres, then rotate them to the $\hat{\bm \beta} \rightarrow \hat{\bm z}$ frame. We then apply these masks to 100 rest frame and boosted 
simulations. We then calculate the power spectra of all maps and look at the relative difference between the boosted and rest frame results in each hemisphere.

\begin{figure}
\centering
Hemispherical Asymmetry
\par \medskip
\text{(a) Temperature }
\par \medskip
\centering
\includegraphics[width=0.99\linewidth]{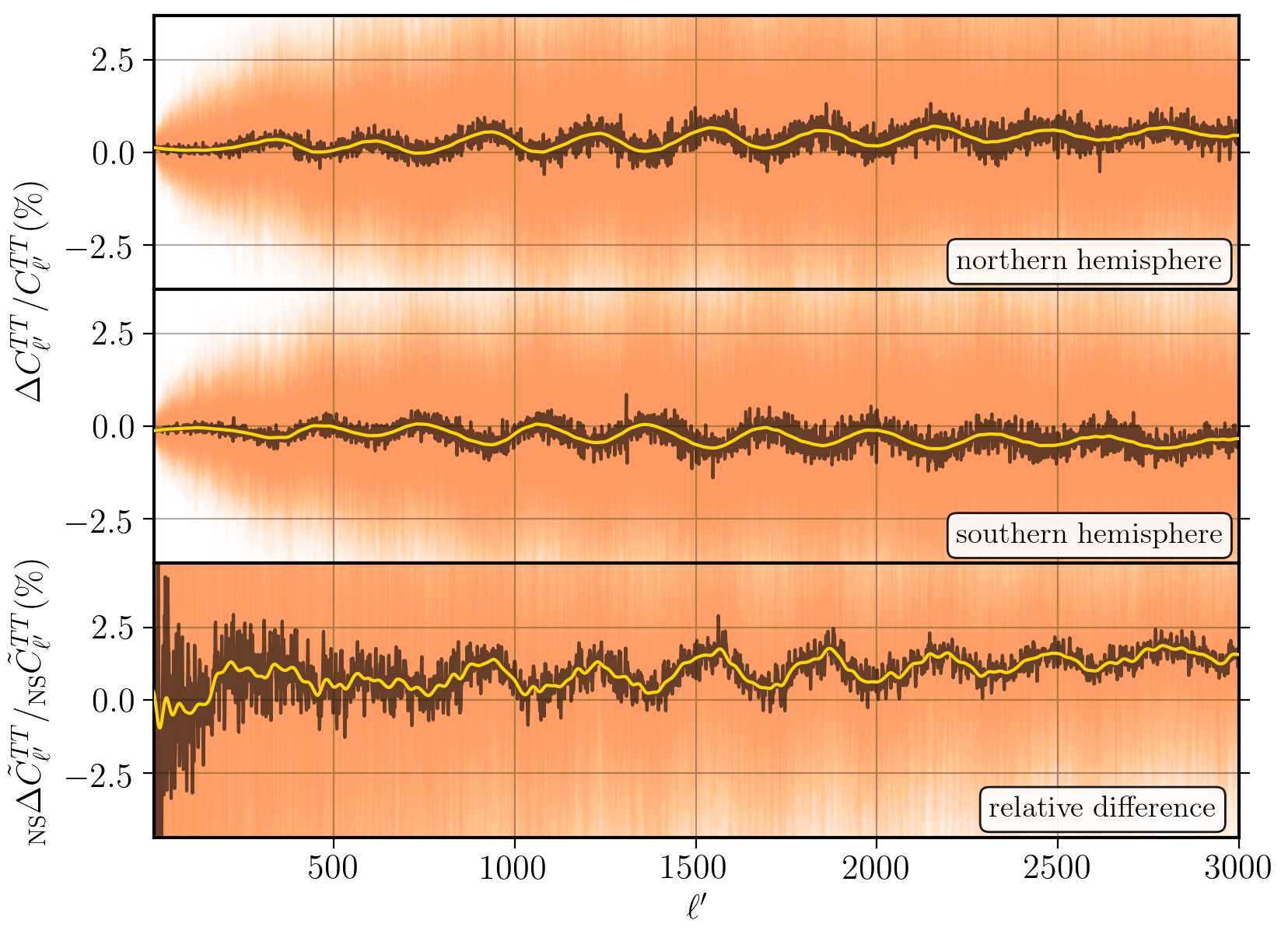}
\text{(b) Polarization}
\par \medskip
\includegraphics[width=0.99\linewidth]{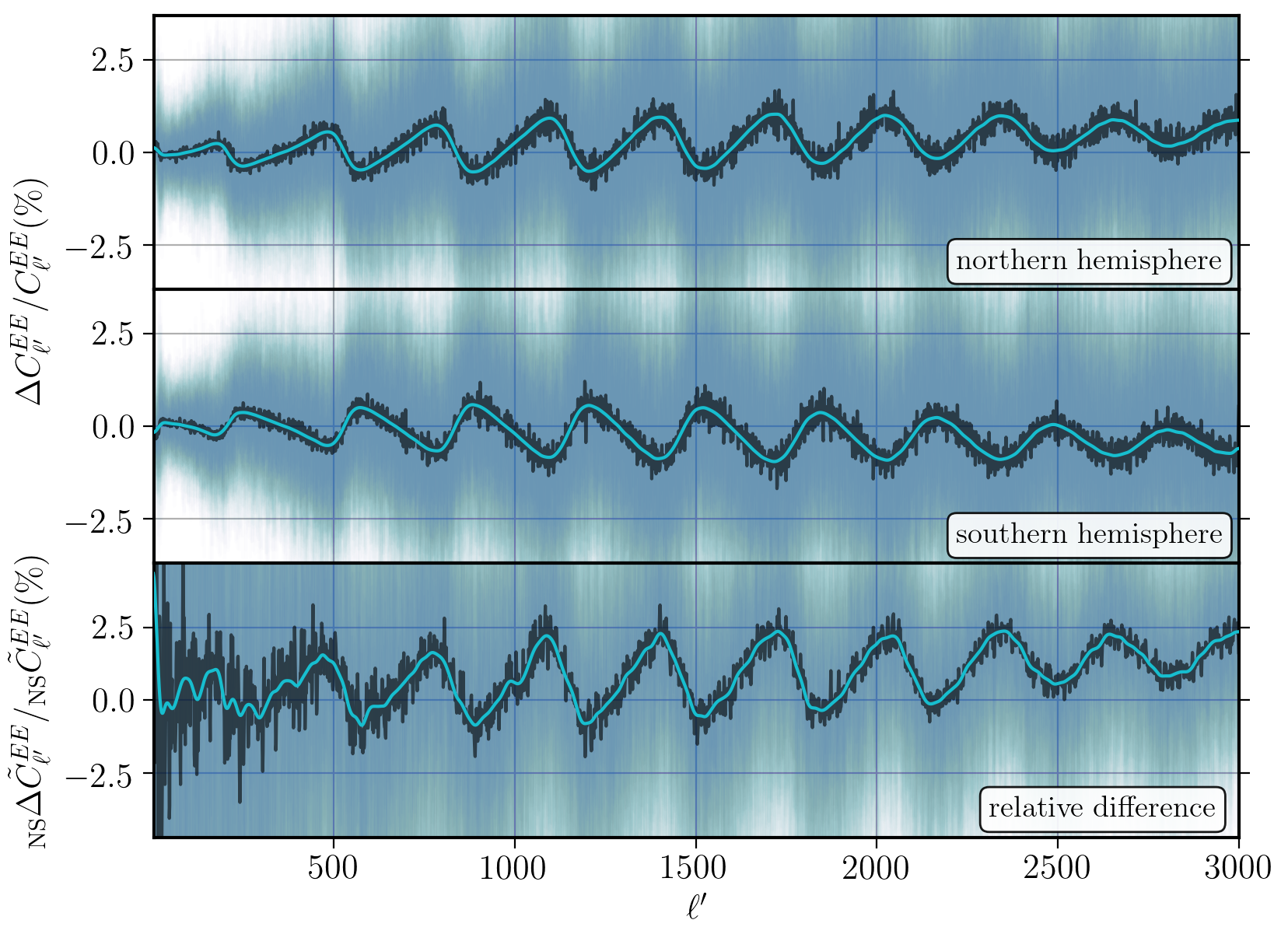}
\centering
\par \medskip
\label{fig:NS_EE_asymmetry}
\caption{Motion-induced hemispherical asymmetry for a Planck-like experiment in (a) temperature and (b) E mode polarization. The \emph{faint lines} in the background (\emph{orange} for TT and  \emph{blue} for EE) are 100 individual simulations plotted on top of each other to show the overall variance of the effect. The \emph{dark jagged lines} are the average of the 100 simulations in all panels. The \emph{colored smooth lines} are the average of simulations Gaussian smoothed over the scale of $\delta \ell' = 10$. As evident from the bottom panels of each plot, Doppler and aberration effects induce percent level power asymmetry between the hemispheres, which is more accentuated in EE. See Eq.~\eqref{eq:NS_delta} in the text for the definition of the statistics presented in the bottom panels.}
\label{fig:NS_asymmetry}
\end{figure}

\begin{figure}
\centering
Hemispherical Asymmetry in Planck 
\par \medskip
\text{(a) Temperature }
\includegraphics[width=0.95\linewidth]{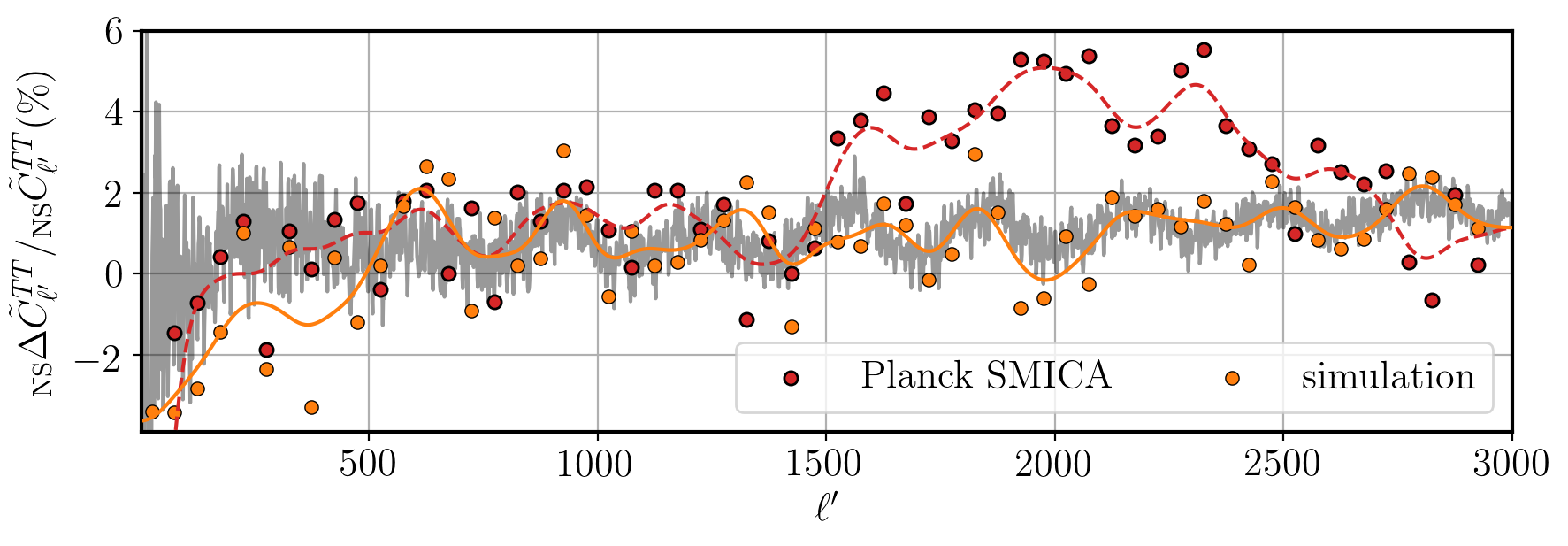}
\text{(b) Polarization}
\par \medskip
\includegraphics[width=0.95\linewidth]{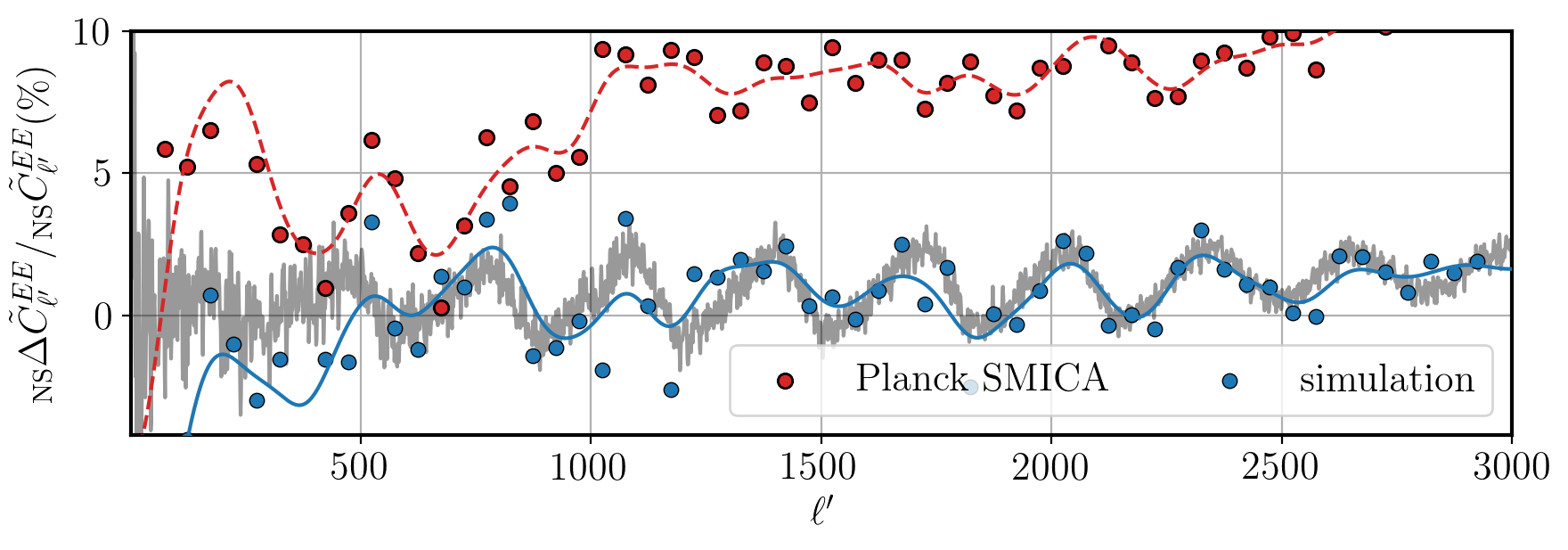}
\caption{Relative difference between the northern and southern hemisphere power spectra in (a) temperature (\emph{top}) and (b) E mode polarization (\emph{bottom}) of the Planck SMICA map binned (\emph{red circles}) and also Gaussian smoothed (\emph{red dashed lines} by $\delta \ell' = 50$. The average of 100 simulations (\emph{grey lines}) are shown here for comparison, as well as an individual simulation binned (\emph{orange and blue circles}) and Gaussian smoothed (\emph{orange and blue lines}). The Gaussian smoothed lines for Planck and simulations are provided for easier visual tracking of the general features of the power asymmetry. The oscillations in north-south difference in power spectra of Planck SMICA marginally follow the ones induced by the boost in simulations, suggesting that they might be partially due to the Doppler and aberration leftovers in the map. See Eq.~\eqref{eq:NS_delta} in the text for the definition of the statistics used in the plots.}
\label{fig:NS_asymmetry_planck}
\end{figure}

Fig. \ref{fig:NS_asymmetry} shows the relative motion-induced power asymmetry in the galactic northern and southern hemispheres for 100 simulations of a Planck-like experiment. The boost is performed assuming the observer is moving in the direction of the CMB dipole with $\beta = 0.00123$ and $\hat{\bm \beta} =  (264^\circ, 48^\circ)$ in galactic coordinates. The average (\emph{jagged lines}) are Gaussian smoothed (\emph{colored lines}) with $\delta \ell' = 10$ for visual guidance, indicating the general behaviour of the oscillations. The top and middle panels of each subfigure depict the relative difference between the power spectrum in the boosted frame $(\tilde{C}^{XX}_{\ell'})$ and rest frame ($C^{XX}_{\ell'}$),
\begin{equation}
    \frac{\Delta C^{XX}_{\ell'}}{C^{XX}_{\ell'}} \equiv \frac{\tilde{C}^{XX}_{\ell'} - C^{XX}_{\ell'}}{C^{XX}_{\ell'}},
\end{equation}
where $X$ is T (\emph{orange}) or E (\emph{blue}). As expected, both TT and EE increase in the northern hemisphere ($\beta \ell \overline{\cos\theta}>0$) and decrease in the southern hemisphere  ($\beta \ell \overline{\cos\theta}<0$). The amplitude of the effect also oscillates according to $\rm{d} C_{\ell} / \rm{d} \ell$, out of phase with the corresponding rest frame power spectra. The bottom panels of each plot show the relative difference between the northern and southern hemispheres

\begin{equation}\label{eq:NS_delta}
     \frac{_{\rm NS}\Delta \tilde{C}^{XX}_{\ell'}}{_{\rm NS}\tilde{C}^{XX}_{\ell'}} \equiv
     2 \frac{_{\rm N}\tilde{C}^{XX}_{\ell'} -~_{\rm S}\tilde{C}^{XX}_{\ell'}}{_{\rm N}\tilde{C}^{XX}_{\ell'} +~_{\rm S}\tilde{C}^{XX}_{\ell'}}.   
\end{equation}
Here, $_{\rm N}\tilde{C}^{XX}_{\ell'}$ and $_{\rm S}\tilde{C}^{XX}_{\ell'}$ represent the boosted power spectra in the northern and southern hemispheres. The advantage of using this statistic to gauge the amount of motion-induced power asymmetry is that it only relies on the boosted frame (observed) power spectra. Therefore, it can be directly applied to observations where the rest frame data are inevitably unavailable. As shown in the bottom panel of both subfigures, the Doppler and aberration effects induce an average of $\sim 1\%$ (with a maximum of $\sim 2.5\%$) increase in TT and EE power spectra. These results are consistent with the one from \citet{Notari:2013} who calculated this effect for temperature (see their Fig. 1).

Now, let us evaluate this statistic on the Planck temperature and polarization data. Fig. \ref{fig:NS_asymmetry_planck} shows the relative difference in power between the northern and southern hemispheres of the Planck SMICA map in comparison with simulations. As evident from the plot, although the hemispherical asymmetry of the SMICA power spectra have the expected sign as the ones induced by the local motion, it is unlikely that the entire deviation is explained by the Doppler and aberration effects. Nevertheless, the oscillations in the north-south power asymmetry of Planck SMICA over various angular scales---especially in the case of TT up to $\ell' \lesssim 1500$---marginally follow the same frequency as the motion-induced oscillations in the simulations, which suggests that these patterns are at least in part induced by the boost. A key point here is that since the Planck maps have not been corrected for the Doppler and aberration effects, we know that their footprints inevitably exist in the in observational data. However, an accurate determination of exactly how much of the effects shown here is due to Doppler and aberration and how much due to foregrounds, mask, and other systematic contaminations requires a more rigorous examination. Here we merely point out the similarity between the motion-induced pattern in observation and simulations, but we avoid over-interpretation of this partial correspondence between the two, and postpone a detailed analysis to future work. A similar plot for the TE power spectrum is presented in Fig. \ref{fig:NS_TE_asymmetry_SMICA}.  

\section{parity asymmetry}\label{sec:parity_asymmetry}

\begin{figure}
\centering
Parity Asymmetry
\par \medskip
\text{(a) Temperature}
\includegraphics[width=0.9\linewidth]{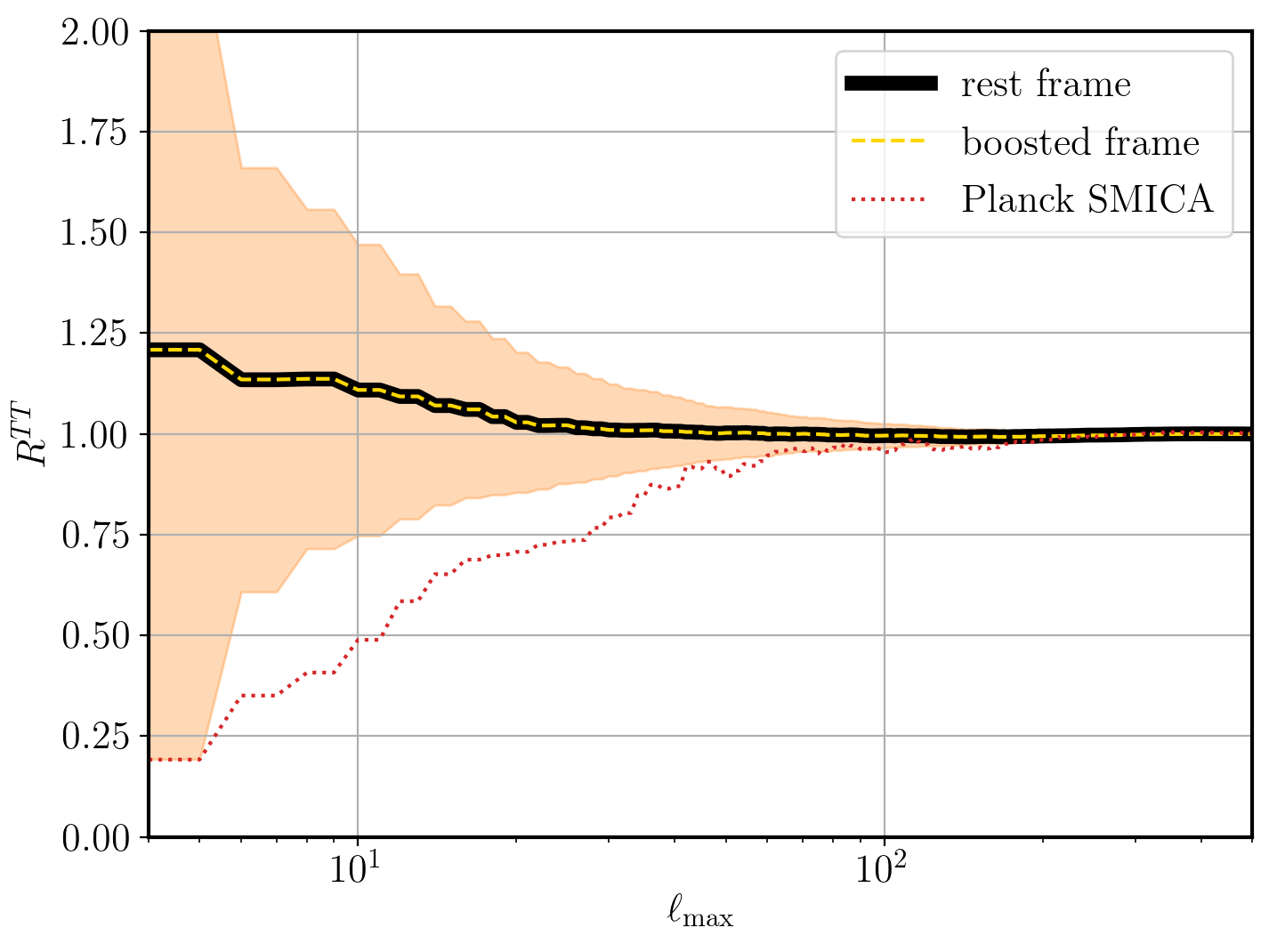}
\par \medskip
\text{(b) Polarization}
\includegraphics[width=0.9\linewidth]{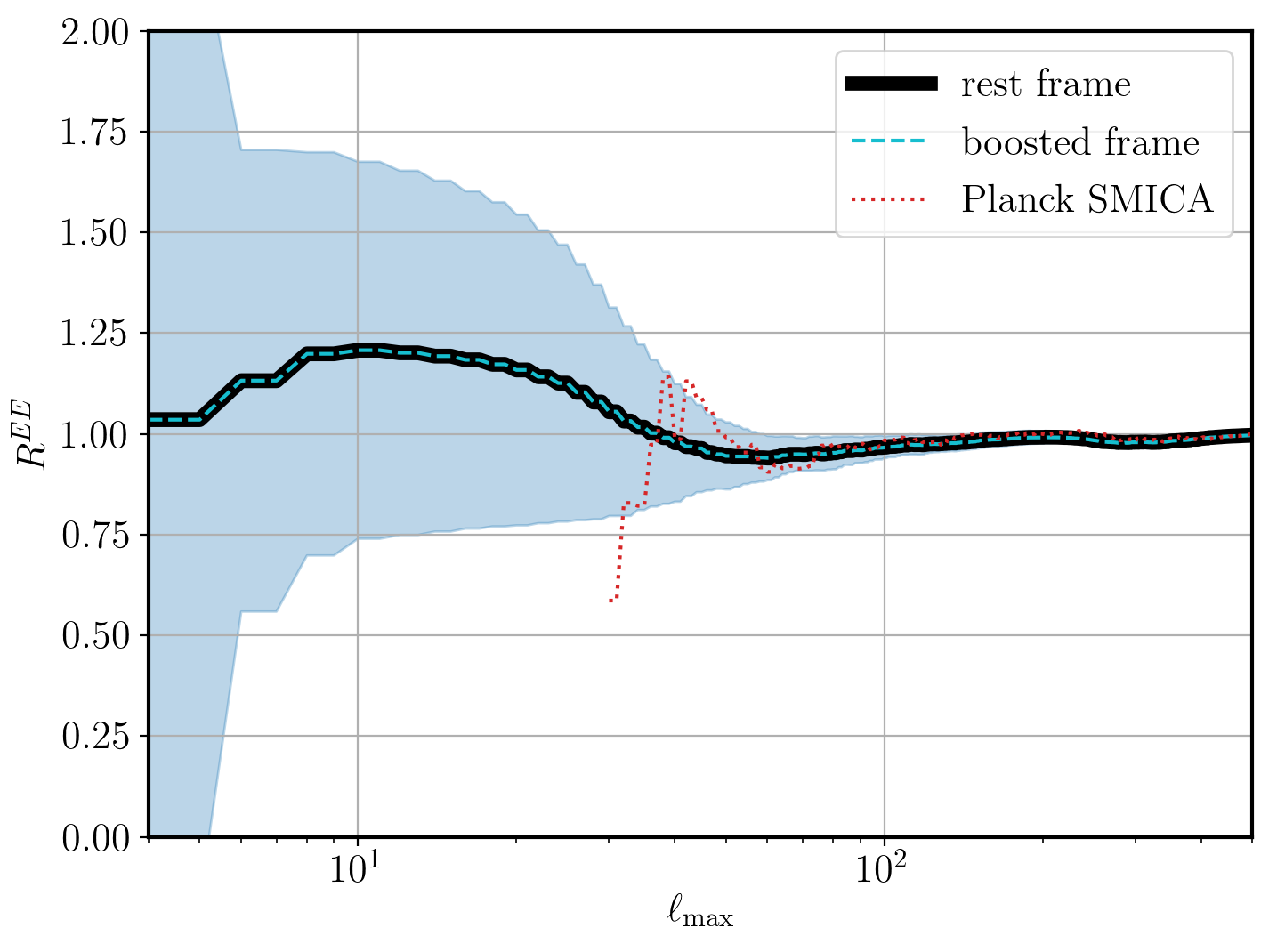}
\caption{Motion-induced parity asymmetry for a Planck-like experiment in (a) temperature \emph{(top)}  and (b) E mode polarization \emph{(bottom)}. The parity asymmetry in the rest frame \emph{(thick solid)} and boosted frame \emph{(dashed)} are almost identical. The shaded regions show the 1-$\sigma$ standard deviation of 100 simulations. The 1-$\sigma$ region for the boosted frame is almost identical to that of the rest frame as well and is not shown in the plot. The parity asymmetry in the Planck SMICA map (\emph{dotted}) is shown for comparison. See Eq.~\eqref{eq:R} in the text for the definition of the statistics presented in the plot.}
\label{fig:parity_asymmetry_planck}
\end{figure}

Another anomaly observed in Planck maps that breaks the expected statistical isotropy is an apparent parity asymmetry: there is more power in the odd multipoles than in the even multipoles \citep{Planck2018:isotropy}. It has been suggested by \citet{Naselsky:2011jp} and \citet{Zhao:2013jya} that the odd-multipole preference of the CMB power spectrum could be induced by the kinematic dipole. In this section, using simulations we examine whether the motion of the observer can cause a parity asymmetry in the CMB temperature and polarization maps. Following section 6.3 of \citet{Planck2018:isotropy}, we use the following statistic to gauge the amount of parity asymmetry in 100 whole-sky temperature and polarization maps

\begin{equation}\label{eq:R}
    R^{XX}(\ell_{\rm max}) \equiv \frac{_{\rm{even}}C^{XX}(\ell_{\rm max}) }{_{\rm{odd}}C^{XX}(\ell_{\rm max}) },
\end{equation}
where 
\begin{equation}\label{eq:Cl_even}
    _{\rm{even}}C^{XX}(\ell_{\rm max}) \equiv \frac{1}{\ell^{\rm even}_{\rm tot}}
    \sum^{\ell_{\rm max}}_{\substack{\ell \in \rm{even}}} \frac{\ell(\ell+1)}{4\pi}C^{XX}_\ell.
\end{equation}
Here $\ell^{\rm even}_{\rm tot}$ is the total number of modes included in the sum. An equivalent definition with \emph{even~$\rightarrow$~odd} holds for the odd multipoles. As before, $X \in \{\rm{T,E}$\}. We apply this statistic to full sky simulations in rest frame and their corresponding boosted equivalents to see if the boost can induce any parity asymmetry in temperature or polarization.

Fig. \ref{fig:parity_asymmetry_planck} shows the average of the parity statistic $R^{XX}$ for 100 simulations in the rest frame and boosted frame, with the shaded areas indicating the 1$\sigma$ regions (standard deviation of the rest frame simulations). The top and bottom plots respectively present $R^{TT}$ and $R^{EE}$. The top panel shows $R^{TT}$ of the Planck SMICA map (\emph{dotted red}), which is about $2\sigma$ away from the simulation average. The average amount of parity asymmetry induced in the simulations up to $\ell_{\rm max}=500$ (range shown in the plot) is only about 0.05$\sigma$ for TT and 0.03$\sigma$ for EE. Therefore, it can be safely concluded that since our local motion does not generate any significant parity asymmetry, it is not the relevant source of this observed anomaly in the CMB. Repeating this exercise only for the north hemisphere (south hemisphere masked) results in $R^{TT} =0.04\sigma$  and $R^{EE}=0.06\sigma$, so it is unlikely that the results will be significantly different in masked skies.

\section{Parameter estimation with Planck power spectrum}\label{sec:param_est}

In this section we investigate how much the motion of the Solar System changes the cosmological parameters inferred from the CMB power spectrum as measured by Planck. As shown in \S~\ref{sec:hemispherical_power_asymmetry}, our local motion changes the power spectrum up to $\sim 2$\% in each hemisphere. However, since Planck's observation patch is relatively symmetric w.r.t. $\betahat$, the effects we observe in the northern hemisphere (contracted angular scales of anisotropies and enhanced amplitudes) get canceled out by the opposite effects in the southern hemisphere (expanded angular scales and diminished amplitudes) when calculating the power spectrum on the whole sky. 

\begin{figure}
\centering
Whole-sky Boosted Simulations 
\par \medskip
\text{(a) Temperature}
\centering
\includegraphics[width=0.9\linewidth]{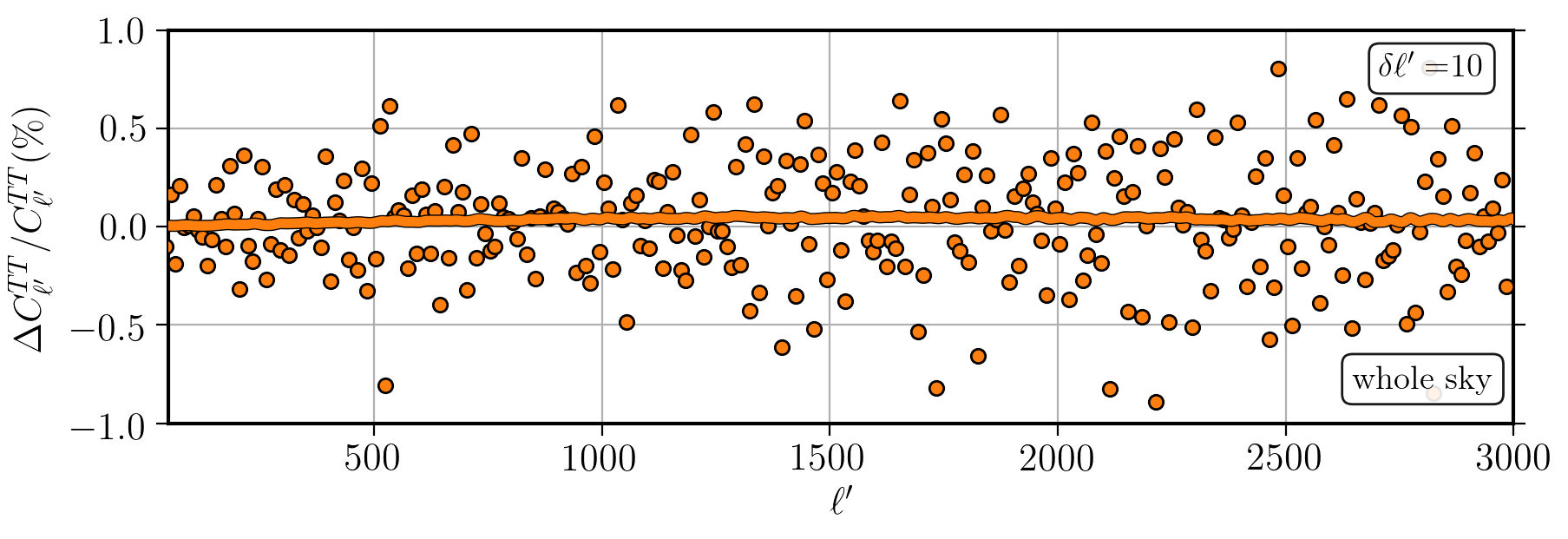}
\text{(b) Polarization}
\par\medskip
\includegraphics[width=0.9\linewidth]{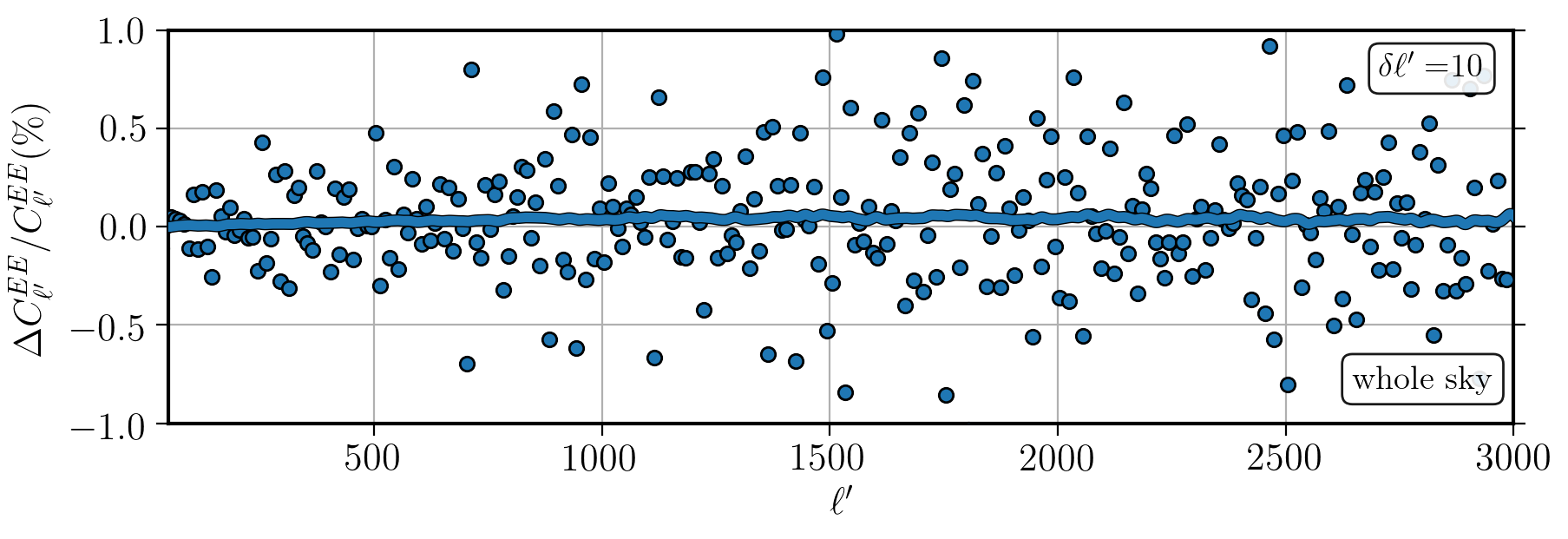}
\centering
\caption{Relative change in the power spectrum of a whole sky map for (a) temperature \emph{(top)}  and (b) E mode polarization \emph{(bottom)}. The average of 100 simulations is Gaussian smoothed with $\delta \ell' =10$ (\emph{solid line}). A random simulation is chosen as a representative of the sample, then binned with $\delta \ell' =10$ \emph{(circles)} in both plots. The overall change on the ensemble average is only about 0.05\% in both plots, but the binned residuals around the average can be as large as 1\%. }
\label{fig:fullsky_Cl}
\end{figure}


Fig.~\ref{fig:fullsky_Cl} shows the effect of the boost on the whole-sky temperature and polarization power spectra. A comparison with Fig.~\ref{fig:NS_asymmetry} reveals how the effects in the northern and southern hemispheres have canceled each other, reducing the effect to only $\mathcal{O}(\beta^2)$ \citep{Burles:2006xf, Challinor2002}. The average change in power spectra reduces to 0.05 \% for both TT and EE\footnote{The fact that $\mathcal{O}(\beta^2)$ effect drops only by a factor of 5 w.r.t. the $\mathcal{O}(\beta)$ effect---not $1/\beta \approx$  800 as naively expected---results from the high degree of non-linearity of the Doppler and aberration kernel.}. Note that similar to the half-sky case, despite the small shift in the average (or Gaussian smoothed) effect, the change in individual modes of $C_\ell$ are still much larger and can reach a few percent. In order to emphasize this fact, we have kept an individual simulation in the plot binned by $\delta \ell' =10$ (the fluctuations on individual $\ell'$ modes  exceed  1\% and are not shown in the plot). The main question here is how much of the change in an individual $C_\ell$ realization propagates to the inferred cosmological parameters in a Planck-like experiment.


In order to assess the relevance of the motion-induced bias for cosmological parameter estimation, we run maximum likelihood analysis on simulated CMB temperature and polarization maps corresponding to the \{100, 143, 217\} [GHz] channels of Planck with the noise configuration \{77.4, 33.0, 46.8\} [$\mu$K-arcmin] for temperature and \{117.6, 70.2, 105.0\} [$\mu$K-arcmin] for polarization and \{10.0, 7.3, 5.0\} [arcmin] beam \citep{Planck2018:overview}. We take the following steps to prepare the dataset for the analysis:

\begin{itemize}
    \item Calculate the TT, EE, and TE theoretical CMB power spectra using the 5 fiducial parameters $(\omega_b, \omega_c, \theta, A_s, n_s)$ (see \S\ref{sec:intro}) with \texttt{CAMB} \footnote{\href{https://github.com/cmbant/CAMB}{\faGithub ~ \texttt{ cmbant/CAMB}}}\citep{Lewis:1999bs}.
    \item Simulate temperature and polarization CMB maps from the theoretical $C_\ell$ using \texttt{healpy.synfast} up to \texttt{lmax=3000} and \texttt{NSIDE=1024}. 
    \item Boost a copy of the simulated map using the \texttt{CosmoBoost} code. This map represents what would be observed in a  frame moving with the velocity $\beta = 0.00123$ in the $\hat{\bm{z}}$ direction. 
    \item Apply the Planck foreground mask to both maps. Since the Doppler and aberration kernel formalism boosts the map in the $\betahat = \hat{\bm z}$ direction, we rotate the mask to the same coordinate system. The output of this step is a CMB map observed in a frame moving with $\beta = 0.00123$ in the $\betahat = (268^\circ, 48^\circ)$ direction.
    \item Calculate the power spectrum of both rest frame and boosted maps using \texttt{pymaster}\footnote{\href{https://github.com/LSSTDESC/NaMaster}{\faGithub ~ \texttt{ LSSTDESC/NaMaster}}}\citep{pymaster} to correct for the coupling among nearby multipoles induced by the mask. 
\end{itemize}

\begin{figure}
	\centering
	\includegraphics[width=1.\linewidth]{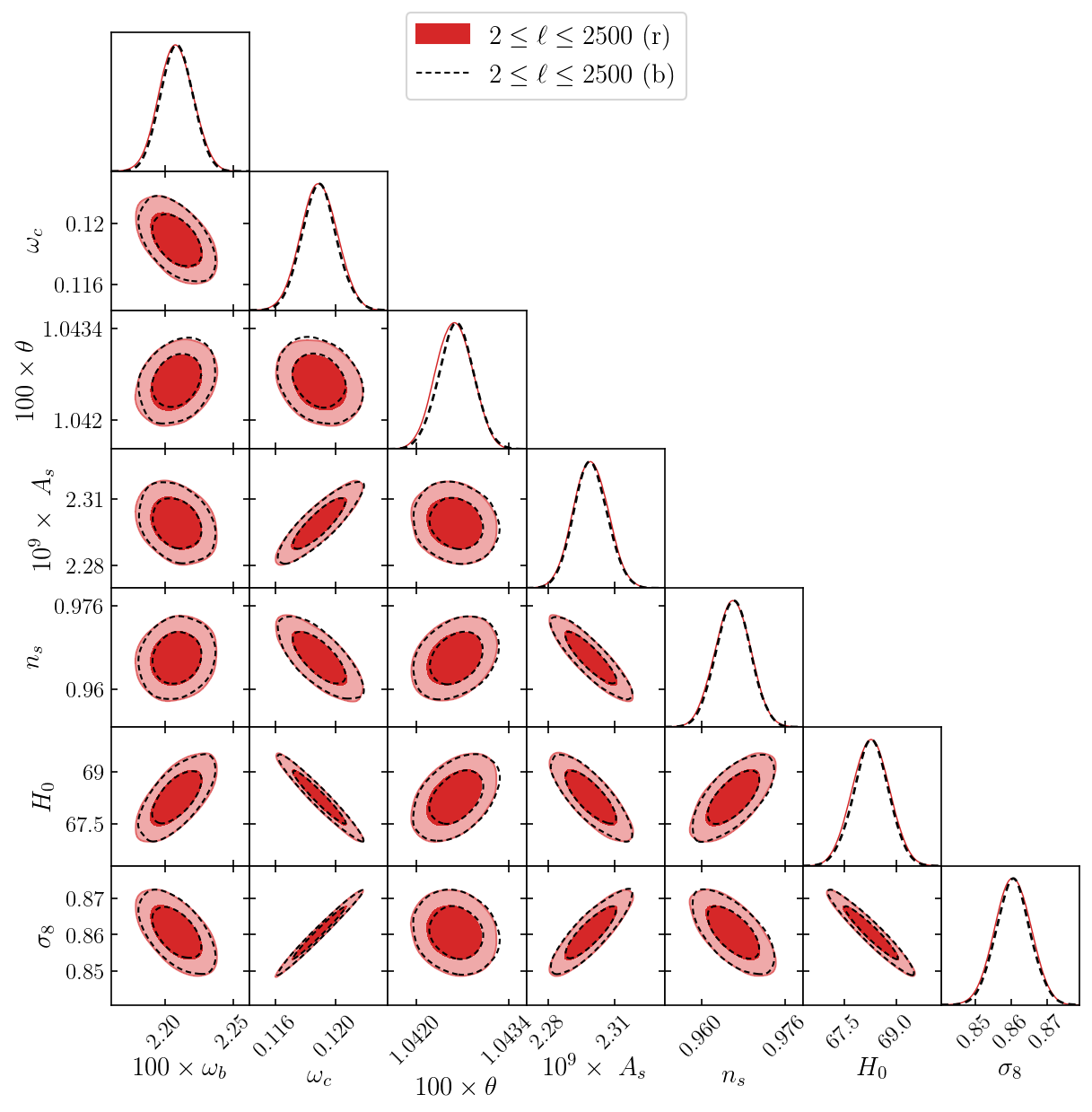}
	\captionof{figure}{Parameter constraints from rest frame and boosted frame TT, EE, TE spectra in the range ($2<\ell<2500$) for a Planck-like experiment. The contour lines are 68\% and 95\% confidence intervals. The shift in parameters are only of the order $0.02\sigma$, which is why the rest frame (\emph{r}) and boosted frame (\emph{b}) contour lines overlap. Here the reionization optical depth $\tau$ is fixed to avoid complications in the low-$\ell$ parameter estimation. See Tab. \ref{tab:planck_7param_all} below for the average shift in parameters due to the boost.}
\label{fig:planck7params_all}

\captionof{table}{The average mean and standard deviation of 10 parameter estimations from boosted and rest frame simulations. $\mu_r$ and $\mu_b$ are the average of the inferred parameters in the rest and boosted frame. $\sigma_r$ is the average variance of each corresponding parameter in the rest frame. The last column reports the shift in the mean of parameters due to Doppler and aberration effects, normalized by the intrinsic error in each parameter. }
\begin{tabular}{ccccc}
& & & & $2<\ell<2500$
\\
Parameter                    & \textbf{$\mu_r$} & \textbf{$\sigma_r$} & \textbf{$\mu_b$} & \textbf{$|\mu_b-\mu_r|/\sigma_r$} \\ \hline\hline
\textbf{$100\times\omega_b$} & 2.199840         & 0.010810            & 2.199801         & 0.033                             \\
\textbf{$\omega_c$}          & 0.118559         & 0.001181            & 0.118573         & 0.021                             \\
\textbf{$100\times\theta$}   & 1.042673         & 0.000156            & 1.042689         & 0.059                             \\
\textbf{$10^9\times A_s$}    & 2.298937         & 0.006144            & 2.298919         & 0.012                             \\
\textbf{$n_s$}               & 0.966851         & 0.002069            & 0.966827         & 0.023                             \\ \hline
\textbf{$H_0$}               & 68.34563         & 0.527770            & 68.34544         & 0.021                             \\
\textbf{$\sigma_8$}          & 0.860056         & 0.005154            & 0.860113         & 0.021                             \\ \hline\hline
\end{tabular}
\label{tab:planck_7param_all}
\end{figure}

\begin{figure}
	\centering
	\includegraphics[width=1.\linewidth]{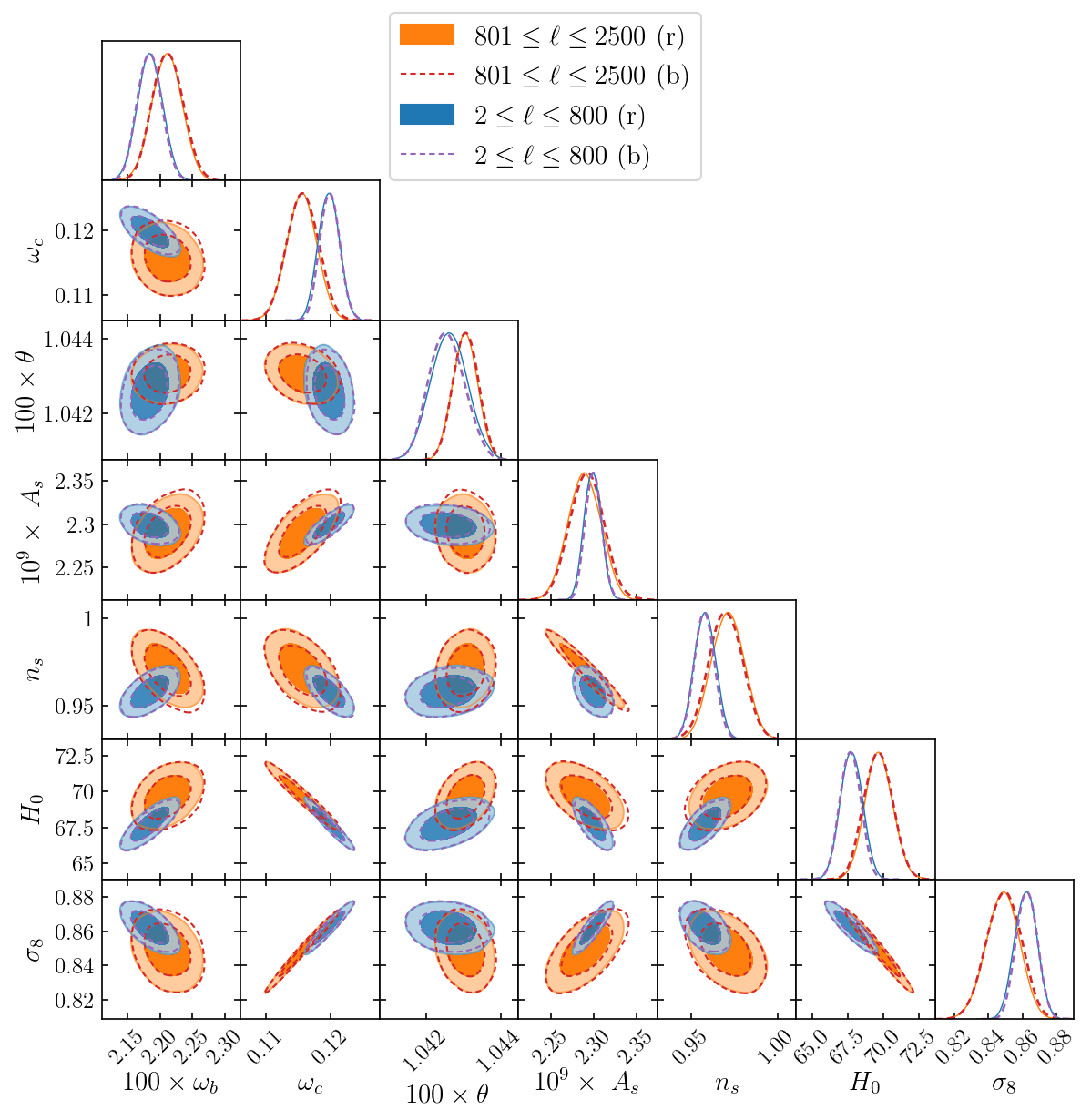}
	\caption{Parameter constraints from rest frame and boosted frame TT, EE, TE spectra in the range $2<\ell<800$ (\emph{blue}) $801<\ell<2500$ (\emph{orange}). The contour lines are 68\% and 95\% confidence intervals. The shift in parameters is slightly larger than the ones in Fig.~\ref{fig:planck7params_all}, but still negligible (see Tab. \ref{tab:planck_7param_lohi}).}
	\label{fig:planck5params_hi-lo}
	
\centering
\captionof{table}{The average motion-induced shift in the mean of estimated parameters in the rest frame ($\mu_r$) and boosted frame ($\mu_b$), normalized by the rest frame uncertainty in each parameter ($\sigma_r$). The results are overall larger than the ones obtained from the whole range of $2<\ell<2500$ in Tab. \ref{tab:planck_7param_all}. }
\begin{tabular}{ccc}
\multicolumn{1}{l}{}         & \multicolumn{1}{l}{$2 \leq \ell \leq 800$} & $801 \leq \ell \leq 2500$         \\
Parameter                    & \textbf{$|\mu_b-\mu_r|/\sigma_r$}          & \textbf{$|\mu_b-\mu_r|/\sigma_r$} \\ \hline \hline
\textbf{$100\times\omega_b$} & 0.042                                      &     0.029                              \\
\textbf{$\omega_c$}          & 0.034                                      &     0.047                            \\
\textbf{$100\times\theta$}   & 0.073                                      &    0.057                               \\
\textbf{$10^9\times A_s$}    & 0.033                                      &      0.060                             \\
\textbf{$n_s$}               & 0.056                                      &      0.059                             \\ \hline
\textbf{$H_0$}               & 0.034                                      &      0.044                            \\
\textbf{$\sigma_8$}          & 0.030                                      &      0.044                            \\ \hline \hline
\end{tabular}
\label{tab:planck_7param_lohi}
\end{figure}

Binning the power spectrum in general reduces the effect of the boost on the change in the power spectrum. Here we do not consider a binning scheme so that we can gauge the maximum possible effect; we use the power spectra with $\delta \ell'_{\rm bin}=1$.

Using the two power spectra obtained from the preparation steps---one in the rest frame, one in the boosted frame---we perform maximum likelihood parameter estimation with the MCMC code \texttt{monte-python 3} \citep{MontePython3}. For simplicity, instrumental noise is directly added to the power spectra, not at the map level. We employ the likelihood of \citet{Hamimeche:2008ai} for two different configurations in TTTEEE : (i) whole range of angular scales probed by Planck ($2 \leq \ell \leq 2500$), and (ii) high $\ell$ ($2 \leq \ell \leq 800$) vs. low $\ell$ ($801 \leq \ell \leq 2500$). 

Fig.~\ref{fig:planck7params_all} shows the results of parameter estimation using TTTEEE spectra in the range $2 \leq \ell \leq 2500$ from one random simulation, both in the rest frame (\emph{red}) and boosted frame (\emph{dotted black}). The contour lines of both frames almost overlap, demonstrating the almost perfect cancellation of the boost effect in the northern and southern hemispheres. To ensure this cancellation is not merely due to chance in the particular simulation we analyzed, we repeat the exercise with 9 more realizations of the power spectra and look at the average shift in the parameters due to boost and compare this with the error bars on each parameter. 

Tab. \ref{tab:planck_7param_all} shows the result of this analysis for all 10 simulations. The average shift on the parameters is on average about $0.02\sigma$, with sound horizon $\theta$ being the most affected ($\sim0.06\sigma$) and curvature perturbation amplitude $A_s$ being the least affected ($\sim0.01\sigma$). The errors on the numbers reported in the last column of the table are less than 20\%, so we decided that 10 simulations suffice for this analysis. 

The results that we find on the shift of parameters are typically smaller than the ones reported by \citet{Catena:2012params} which only considers temperature. This is likely due to the fact that their boosting formalism slightly overestimates the effect (see Fig. 2 in \citet{Dai2014}). Inclusion of polarization information (EE, TE) does not affect average shift in parameters significantly compared to chains that only run with the TT power spectrum. The EE-only runs also yield similarly negligible shifts in parameters.

Fig. \ref{fig:planck5params_hi-lo} shows similar results for the TTTEEE parameter estimation but with ($2 \leq \ell \leq 800$) and ($801 \leq \ell \leq 2500$) plotted separately. In the $2 \leq \ell \leq 800$ regime, $\theta$ is still the most affected parameter---understandably so, because it is primarily determined by the position of the first peak. However, in the $801 \leq \ell \leq 2500$ range the two parameters $A_s$ and $n_s$ are also affected comparably. Overall, the shift in parameters is still negligible and remains smaller than $0.1 \sigma$ for all of them.  In this analysis we did not consider the frequency dependence of the boost effect which is negligible for experiments with symmetric (w.r.t. the direction of motion) masks such as Planck \citep{Yasini:2016pby}. This effect could be potentially important for partial sky experiments but we postpone a detailed analysis of its implications for cosmology to future studies.

\section{Boost detection using harmonic mode coupling}\label{sec:boost_detection}

\begin{figure*}
\centering
\begin{tabular}{c  c  c}
    \includegraphics[width=.32\textwidth]{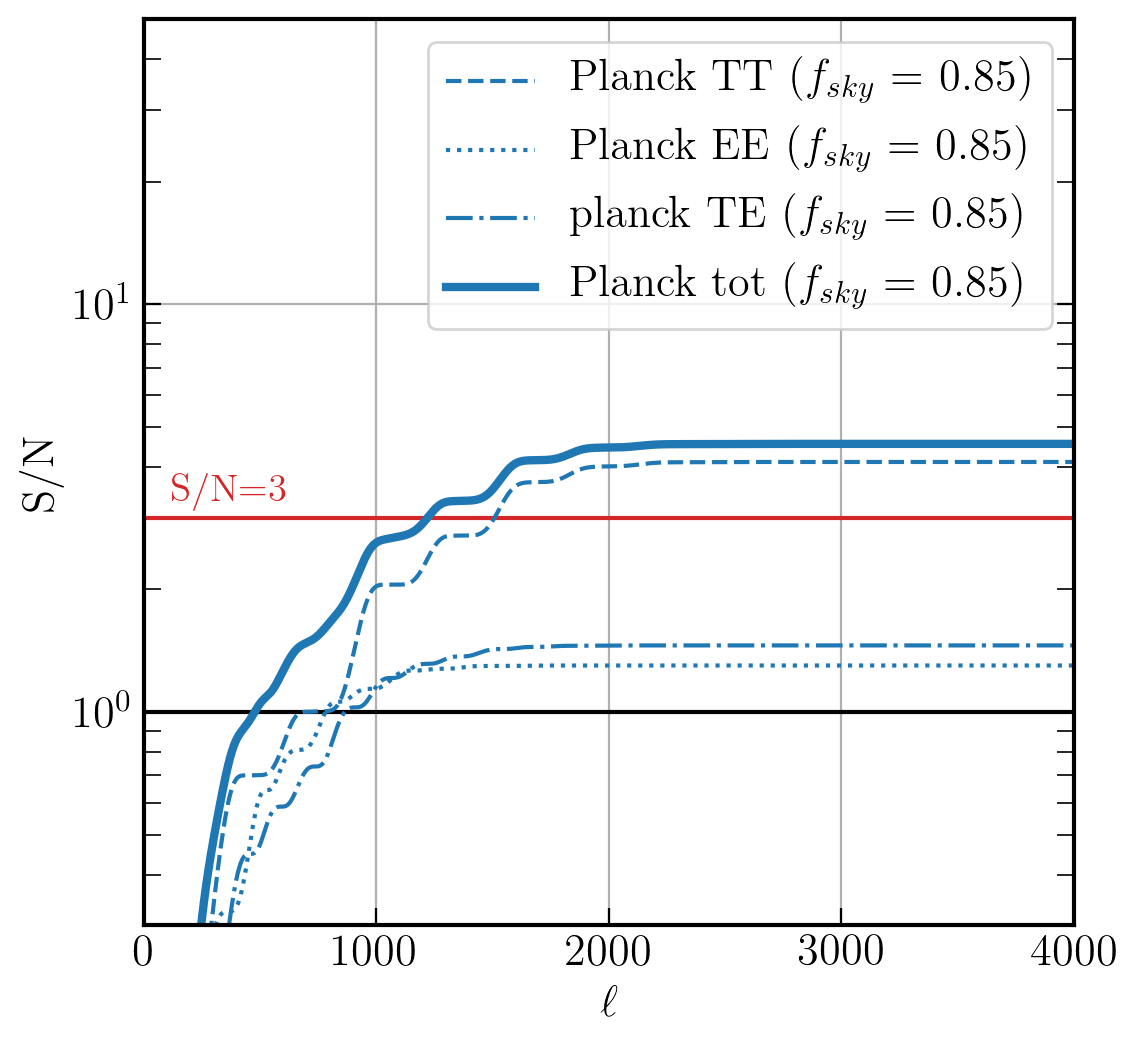} &
    \includegraphics[width=.32\textwidth]{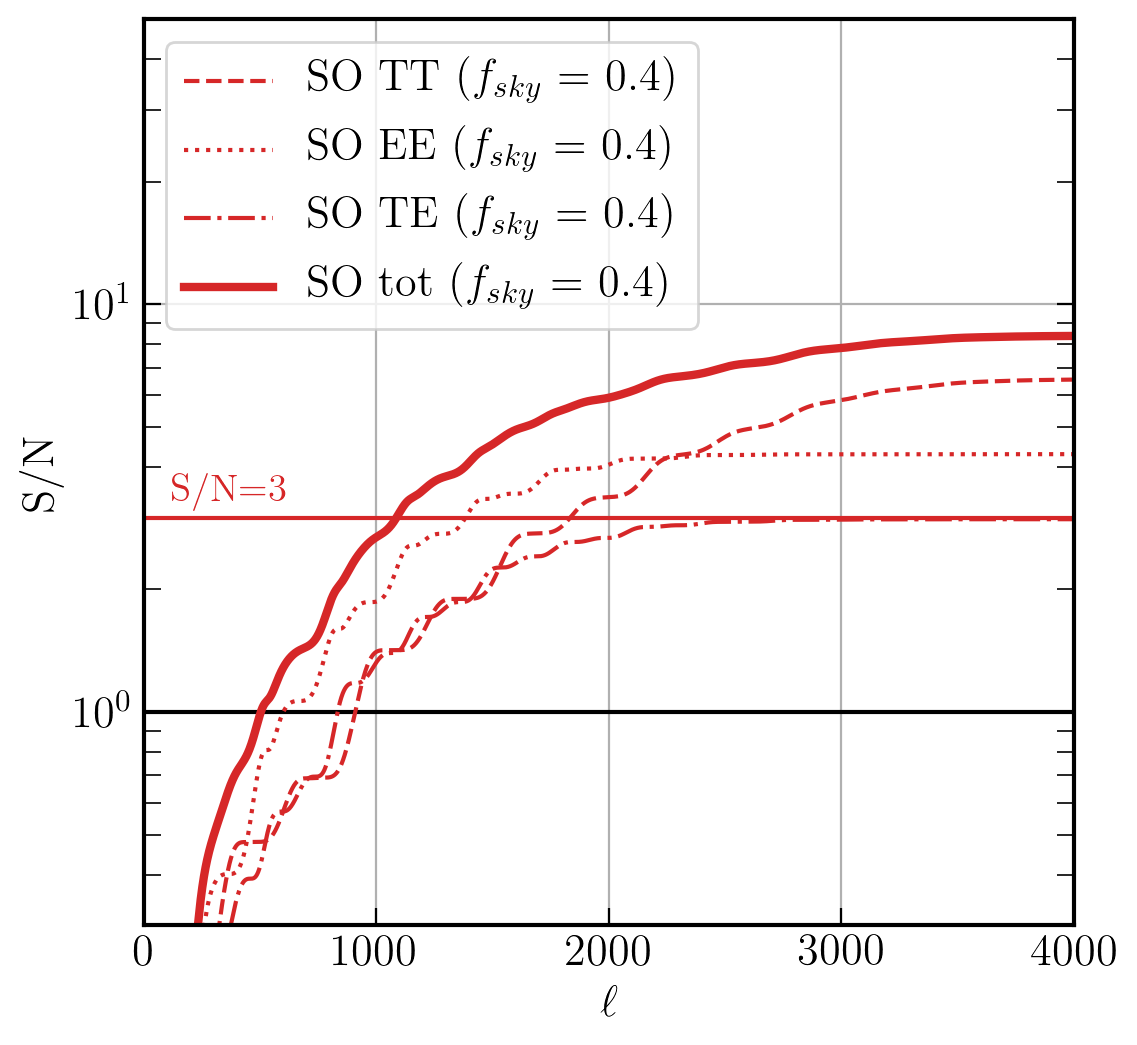} &
    \includegraphics[width=.32\textwidth]{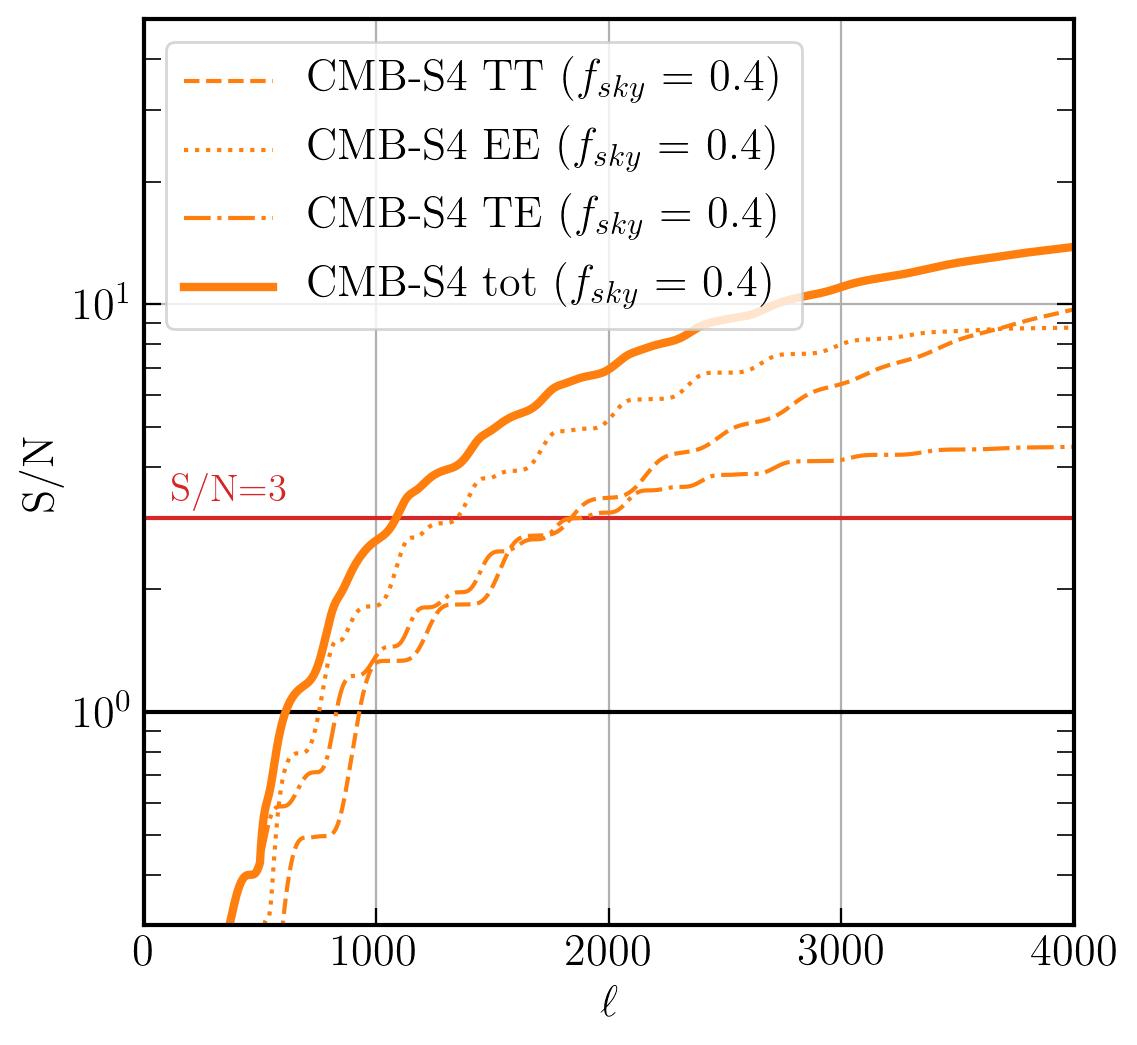}\\
    \small (a) Planck & 
    \small (b) Simons Observatory &
    \small (c) CMB-S4
  \end{tabular}
\caption{Boost detection (via mode coupling) signal-to-noise estimates for (a) Planck (\emph{left}), (b) Simons Observatory (\emph{middle}), and (c) CMB-S4 (\emph{right}). The maximum achievable S/N for these experiments up to $\ell_{\rm max}=4000$ are respectively $\sim4$, $\sim8.5$, and $\sim20$. }
\label{fig:boost_S/N}
\end{figure*}

The amount of motion-induced leakage of the nearby multipoles into each other depends on the amplitude and direction of the local peculiar velocity vector with respect to the CMB rest frame. Therefore, measurements of this induced mode coupling in CMB temperature and polarization---which have well-known statistics---can be exploited to measure the local peculiar velocity vector. This idea has been investigated in detail in \citet{Kosowsky2010,Amendola2010, Notari:2011predeboost}, and the signal has been detected by Planck using $500<\ell<2000$ of 143 and 217 GHz channel maps \citep{Aghanim:2013suk} reporting a value of $\beta=0.00128 \pm 0.00026(\rm{stat.})\pm0.00038(\rm{syst.})$. Here we present estimates for the achievable S/N with future surveys such as SO and CMB-S4, and compare their performance with Planck. 

The amplitude of the motion-induced coupling between mode $\ell'$ and and its first neighbor $\ell'+1$ in the moving frame can be calculated using Eq.~\eqref{eq:alm_kernel} as the following \citep{Amendola2010, Chluba2011, Dai2014}

\begin{align}\label{eq:Flm_boost_signal}
    \langle F_{\ell' m}\rangle \equiv& \langle \tilde{a}^{X*}_{\ell' m} \tilde{a}^Y_{(\ell'+1) m}\rangle\nonumber\\
    =& \sum_\ell~\Ker{X}{\ell'}{\ell}(\beta) ~\Ker{Y}{(\ell'+1)}{\ell}(\beta) C^{XY}_\ell\nonumber\\
    \overset{\overset{1/\beta \gg \ell' \gg 1}{\downarrow}}{\simeq}&
    \beta B^Y_{(\ell'+1)m} C^{XY}_{\ell'}
    -\beta B^X_{(\ell'+1)m} C^{XY}_{\ell'+1},
\end{align}
with
\begin{align}
    B^{X}_{\ell m} \equiv \sqrt{\frac{(\ell^2-m^2)(\ell^2-s_{X}^2)}{(4\ell^2-1)}} \hspace{2em} \& \hspace{2em} X \rightarrow Y
\end{align}
where $X,Y \in \{\rm{T,E}\}$ in thermodynamic temperature units (Doppler weight 1), and $s_X$ and $s_Y$ are the spin weights of the observable (0 for temperature and 2 for polarization). The approximation in the last line of Eq.~\eqref{eq:Flm_boost_signal}, presented in \citet{Amendola2010} is not valid for $\ell'\gtrsim 1/\beta \approx 800$ and overestimates the amount of coupling between first neighbors ($\Delta \ell'=1$). Nevertheless, as pointed out in \citet{Chluba2011} this formula approximates the total signal due to the relevant coupling between all the nearby neighbors very well\footnote{We checked with simulations and found that up to $\ell_{\rm max} =3000$, the formula captures the contribution of the first 4 neighbors with 0.1\% accuracy.}. Therefore, to avoid numerical complications we use this approximation as a proxy to assess the attainable signal-to-noise for \emph{boost detection} with Planck, SO, and CMB-S4.  We also use the estimate for the covariance\footnote{The accuracy of this estimation was not checked with simulations.} from \citet{Amendola2010}
\begin{equation}\label{eq:Flm_boost_noise}
    \sigma^2_{\ell' m} \approx \mathfrak{C}_{\ell'} \mathfrak{C}_{\ell'+1}.
\end{equation}
Here $\mathfrak{C}_\ell\equiv (C_\ell + N_\ell)/\sqrt{f_{\rm sky}}$ is the effective power spectrum and $N_\ell \equiv w^{-1}\exp[{\ell(\ell+1)\theta_{\rm fwhm}^2/(8\log{2})}]$ is the instrumental noise, where $w^{-1}$ is the sensitivity [$\mu$K-arcmin] and $\theta_{\rm fwhm}$ is the full-width-half-maximum of the beam [arcmin]. Using Eqs. \ref{eq:Flm_boost_signal} and \ref{eq:Flm_boost_noise} we can easily calculate the total signal-to-noise for different experiments using their \{$w^{-1}, \theta_{\rm fwhm}, f_{\rm sky}$\} configurations as 

\begin{align}
\frac{S}{N} = 
\sqrt{
    \sum_{\ell',m}
    \frac{
        \left( \beta B^Y_{(\ell'+1)m} C^{XY}_{\ell'}
        -\beta B^X_{(\ell'+1)m} C^{XY}_{\ell'+1}\right)^2
        }{
        \mathfrak{C}_{\ell'} \mathfrak{C}_{\ell'+1}  
    } 
}.
\end{align}
Using this equation, we now estimate the S/N for measuring $\beta$ (assuming $\betahat$ is known) with Planck, SO and CMB-S4. For Planck ($f_{\rm sky} = 0.85$), we use the combined 143 and 217 GHz noise power spectra to emulate the results of \citet{Aghanim:2013suk} which uses the data from these two frequency channels. The sensitivity and beam size for these channels are shown in Tab. \ref{tab:survey_noise}). For SO ($f_{\rm sky} = 0.4$) we use the closest matching frequencies 145 and 225 GHz (see Tab. \ref{tab:survey_noise}), and for CMB-S ($f_{\rm sky} = 0.4$) we use the nominal $w_T^{-1}=1~\mu$K-arcmin and $\theta_{\rm fwhm}=1.4$ arcmin.

\begin{table}
\caption{Sensitivity and beam configuration used for Planck, Simons Observatory, and CMB-S4.}
\begin{tabular}{lccc}
                          & \textbf{Temperature} & \textbf{Polarization} & \textbf{Beam} \\
                           & {[$\mu$K-arcmin]} & {[$\mu$K-arcmin]} & {[arcmin]} \\
                           \hline\hline 
\textbf{Planck (143 GHz)} & 33       & 70        & 7.3       \\
\textbf{Planck (217 GHz)} & 47       & 105        & 5.0       \\
\textbf{SO (145 GHz)}     & 10       & 14        & 1.4       \\
\textbf{SO (225 GHz)}     & 22       & 31        & 1.0       \\
\textbf{CMB-S4 nominal}   & 1      & 1.4        & 1.4      
\end{tabular}
\label{tab:survey_noise}
\end{table}

Fig. \ref{fig:boost_S/N} shows the estimated S/N from Planck (\emph{left}), Simons Observatory (\emph{middle}), and CMB-S4 (\emph{right}). Each plot shows the S/N in TT (\emph{dashed}), EE (\emph{dotted}), and TE (\emph{dot-dashed}) separately, and all combined (\emph{solid}). Planck clearly becomes noise dominated at $\ell>2000$, and it can only obtain a S/N of about 4.5 with minimal contribution from EE and TE. The Simons Observatory, despite its smaller sky fraction and hence coverage of lower number of $m$ modes for a given $\ell$, has a much lower instrumental noise and can supersede Planck in detection of this signal. Using only EE, SO can achieve a similar S/N to Planck (compare the \emph{dotted line} in the middle panel to the \emph{solid line} in the left). Combining TT, EE, and TE for SO results in a total S/N of 8.5, roughly twice as large as Planck's. Notice that below $\ell \sim 2000$ where the SO EE is not noise limited, it gains a large S/N compared to TT (compare \emph{dotted} and \emph{dashed lines} in the middle panel). As pointed out before, since the EE power spectrum fluctuates more abruptly than TT, it yields a larger mode coupling in the boosted frame. Finally, for comparison we also show the results for CMB-S4 which can attain a total S/N of around 20. 

\begin{figure}[b]
    \centering
    \includegraphics[width=0.35\textwidth]{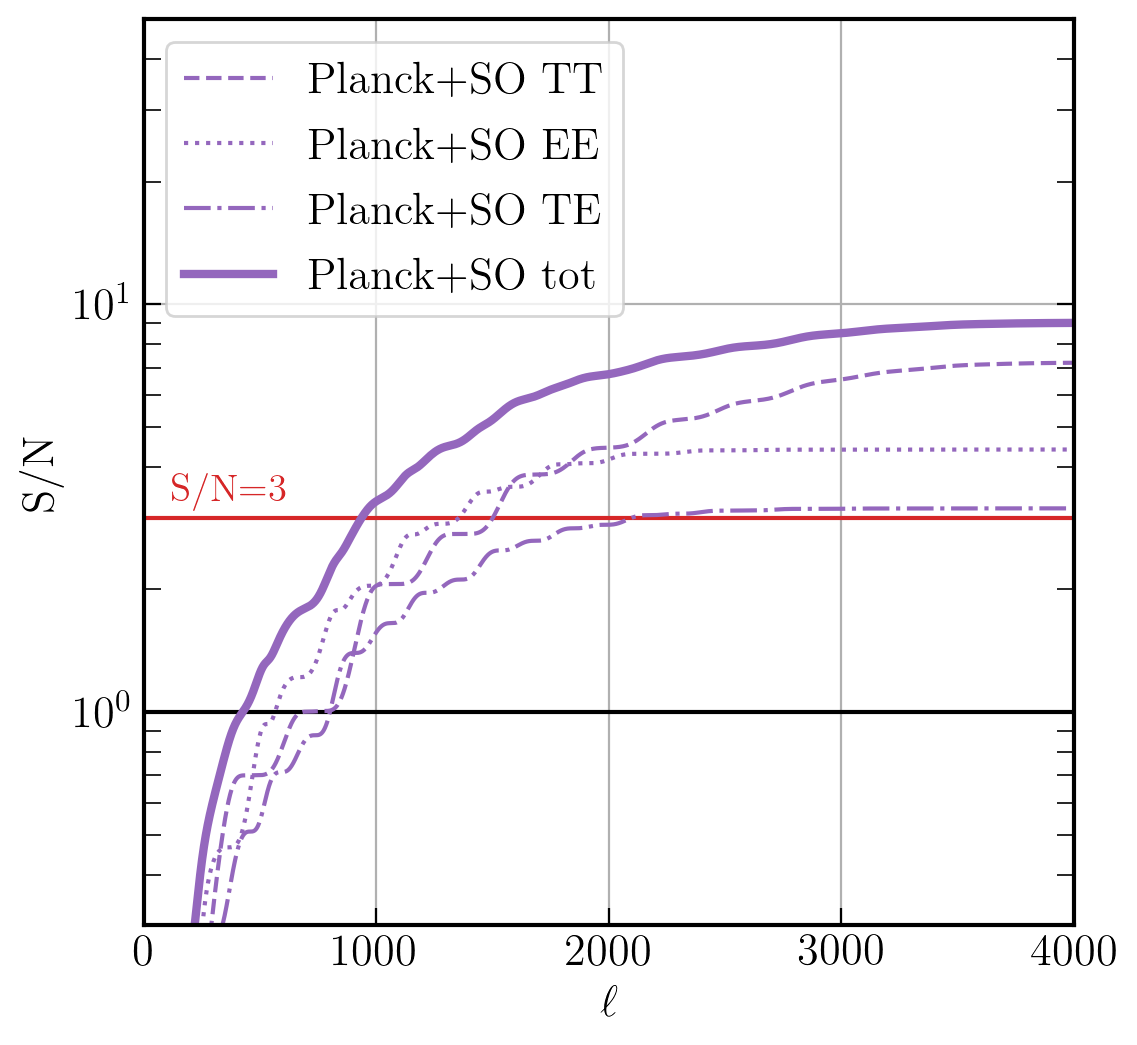}
    \caption{Combining Planck and SO can increase the total attainable S/N for boost detection to $\sim9$. }
    \label{fig:SO+Planck}
\end{figure}
This detection leverages on mode-coupling to measure $\beta$ and is highly sensitive to number of available modes in the observed patch and hence $f_{\rm sky}$. Since Planck covers a larger sky fraction compared to SO and CMB-S4, it is interesting to see whether there are any synergies between these experiments. To assess this, we combine the signal from the additional sky fraction covered by Planck, and add it to the obtained signal from SO, and then separately to CMB-S4. In order to find the effective noise in the common sky areas covered by different experiments, we find the harmonic mean of their individual noise power spectra squared. Fig. \ref{fig:SO+Planck} shows the combination of Planck+SO. As we can see, the contribution to the total S/N by SO only improves from 8.5 to 9. For CMB-S4 the improvement is virtually zero and the result is identical to the right plot in Fig.~\ref{fig:boost_S/N}.

\section{Summary and Discussion}\label{sec:summary}

In this paper we reviewed how the motion of the observer affects the statistics of the cosmic microwave background radiation, expanding previous studies in various ways. We  adopted a  more careful boosting methodology which is also applied to polarization maps and spectra, we clarified how deboosting can and should be performed, and we assessed how boosting effects could be exploited in future experiments to measure our peculiar motion with respect to the CMB.
This paper also introduces 
\textbf{CosmoBoost} \href{https://github.com/syasini/CosmoBoost}{\faGithub}: a software aimed at  accurately simulating  the Doppler and aberration effects, or correcting/removing their footprints from observations.

We investigated the significance of the Doppler and aberration effects---induced by a dimensionless velocity  $\beta=0.00123$ towards $\betahat=(264^\circ, 48^\circ)$---on CMB statistics. The summary of the main  results is as follows:

\textbf{Boost Variance:} The motion of the observer changes the variance of the CMB anisotropies around the theoretical mean. Using simulations, in \S \ref{sec:boost_variance} we showed that the boost variance in an area towards the direction of motion ($f_{\rm sky} = 10 \%$) becomes larger than 10\% of the rest frame cosmic variance in temperature, and 20\% in polarization beyond $\ell'\gtrsim 1500$. However, if the effect of the boost is being corrected in the theoretical mean, the boost variance reduces by an order of magnitude. We showed that the correction to the mean does not convey all the information on $C_\ell$ modes in an individual (boosted) observation. We suggested the use of CosmoBoost to correct boosting effects on the spherical harmonics, and we provided practical advice on how to perform the correction to the CMB power spectrum without using this software. 

\textbf{Hemispherical Power Asymmetry:} In small patches of the sky towards (opposite to) the direction of motion, the induced coupling between harmonic multipoles increases (decreases) the power in both temperature and polarization up to a few percent. In \S\ref{sec:hemispherical_power_asymmetry} we showed that this difference in power results in a hemispherical asymmetry in Planck-like simulations with a average (maximum) of roughly 1.0\% (1.7\%) in TT, and 0.9\% (2.1\%) in EE (see Fig. \ref{fig:NS_asymmetry}). The motion-induced distortions in the power spectra ($\propto \diff \C_{\ell}/\C_\ell$) are out of phase with the primary CMB fluctuations, therefore, any hemispherical asymmetry of this origin is expected to show a similar fluctuating pattern. We made a comparison between the scale dependent fluctuations in the power asymmetry of the Planck SMICA spectra and the ones from boosted simulations. The marginal resemblance between the oscillating patterns in the  observation and simulations suggests that part of the hamispherical asymmetry observed in Planck SMICA maps could be due to the Doppler and aberration effects. However, we do not draw final conclusions and postpone a more rigorous analysis of this effect to future work.

\textbf{Parity Asymmetry:} An analysis of the whole-sky boosted and rest frame simulations revealed that the motion of the observer does not induce any parity asymmetry in neither temperature nor polarization of the CMB. In \S~\ref{sec:parity_asymmetry}, using the $R$ statistic which calculates the ratio of even ($\ell=2n$) to odd multipoles $(\ell=2n+1)$, we only found a motion-induced parity asymmetry of the order of a few percent of the standard deviation in rest frame simulations  in  both TT and  EE, for $\ell_{\rm max}=500$. Therefore, we safely reject Lorentz boost as a source of the anomalous parity asymmetry observed in Planck maps.

\textbf{Planck Cosmological Parameter Estimation:} In \S~\ref{sec:param_est} we revisited the impact of the boost on the Planck cosmological parameter estimation in light of more precise boosting scheme, accurate mask consideration and inclusion of polarization spectra.
The bias implied  by a neglected boost correction  is only about $0.02\sigma$ on most parameters, with the most (least) affected being the sound horizon $\theta$ at $0.06\sigma$ (the amplitude of curvature perturbation $A_s$ at $0.01\sigma$).
Therefore, the inclusion of these  sophistication in the analysis
shows  even  smaller biases on parameters  than the ones reported in \citet{Catena:2012params}.

\textbf{Boost Detection:}  
The motion-induced coupling between nearby harmonic multipoles can be exploited to measure our peculiar velocity with respect to the CMB rest frame in a different way than observing the CMB dipole.
While Planck has measured our peculiar velocity using only TT spectra up to $l \simeq 2000$ (\citet{Aghanim:2013suk}), future experiments like SO and CMB-S4 could also leverage on polarization information (albeit on a smaller area of the sky).
The signal, which is proportional to the slope of the underlying power spectrum, is intrinsically larger in EE polarization than TT. In \S \ref{sec:boost_detection} we showed that exploiting the high temperature and polarization sensitivity of the Simons Observatory will allow to obtain $S/N \sim 8.5$, and $S/N \sim 9$ in combination with Planck. CMB-S4 can enhance this measurement to $S/N \sim 20$. These measurements combined with other probes of the the local motion of the Solar System such as galaxy number counts \citep{Pant:2018smd} and SZ clusters \citep{Chluba:2004vz} will ultimately yield a Dipole independent measure of our motion w.r.t. the CMB.


\appendix

\section*{Acknowledgements}

We are sincerely grateful to Daniel Lenz, Thejs Brinckmann, Lukas Hergt, Antony Lewis, and Nareg Mirzatuny for their incredible help with technical aspects of the project. We thank Loris Colombo, Emmanuel Schaan, and Colin Hill for helpful comments and discussions. We are also extremely thankful to the anonymous referee for their helpful comments on our manuscript. EP is supported by NASA 80NSSC18K0403 and the Simons Foundation award number 615662; EP and SY are supported by NSF AST-1910678. We acknowledge the use of the following softwares: \texttt{CAMB} \citep{Lewis:1999bs}, \texttt{Healpix} \citep{Healpix}, \texttt{NaMaster} \cite{pymaster}, \texttt{MontePython 3} \citep{MontePython3, MontePython}, and \texttt{GetDist} (\href{https://github.com/cmbant/GetDist}{\faGithub \texttt{ cmbant/getdist}}). Computations for this work were supported by the Center for High-Performance Computing (HPC) at University of Southern California.




\bibliographystyle{mnras}
\bibliography{cosmobib}





\appendix

\section{Doppler and aberration kernel}\label{sec:app:kernel}

For reference, here we explicitly write the expressions for the temperature and polarization Doppler and aberration kernels. Detailed discussions on the topic can be found in \citet{Challinor:1999yz,Chluba2011,Dai2014,Yasini:2017jqg}.

We start with the boost equation in real space 
\begin{equation}
    \tilde{T}(\gammahat') = \frac{T(\gammahat)}{\gamma(1-\beta \betahat \cdot \gammahat')},
\end{equation}
and expand the temperature on both sides in terms of spherical harmonics 
\begin{equation}
    \sum_{\ell'm'}\tilde{a}^{T}_{\ell' m'}Y_{\ell' m'}(\gammahat') = \frac{1}{\gamma(1-\beta\mu')} \sum_{\ell m}a^{T}_{\ell m}Y_{\ell m}(\gammahat).
\end{equation}
Now in order to isolate $\tilde{a}^{T}_{\ell' m'}$ on the left hand side, we use the orthogonality condition of spherical harmonics and multiply by $Y^*_{\ell' m'}(\gammahat')$, then integrate over the all angles to get 
\begin{equation}
    \tilde{a}^{T}_{\ell' m'}=  \sum_{\ell m} \int {\diff^2} \gammahat' \frac{Y^*_{\ell' m'}(\gammahat')Y_{\ell m}(\gammahat) }{\gamma(1-\beta\betahat \cdot \gammahat')} a^{T}_{\ell m}.
\end{equation}
which can be simplified as 

 \begin{equation}
 \tilde{a}^T_{\ell' m} =
 \sum_{\ell m}~_T\mathcal{K}^{m' m}_{\ell' \ell}(\beta) ~a^T_{\ell m}.
 \end{equation}
In the $\betahat = \hat{\bm z} $ coordinate system, the integral yields a $\delta_{m' m}$, implying that there is no mixing in $m$. Therefore, the $m'$ index of the kernel is redundant and can be dropped. This simplifies the definition of the Doppler and aberration kernel for temperature to 

\begin{equation}
    \Ker{T}{\ell'}{\ell}(\beta)=  \int {\diff^2} \gammahat' \frac{Y^*_{\ell' m}(\gammahat')Y_{\ell m}(\gammahat) }{\gamma(1-\beta  \hat{\bm z} \cdot \gammahat')}.
\end{equation}
We can easily derive the equivalent expressions for polarization E and B modes as \citep{Dai2014,Yasini:2017jqg}

\begin{equation}
    \Ker{E}{\ell'}{\ell}(\beta)=  \int {\diff^2} \gammahat' \frac{_2Y^*_{\ell' m}(\gammahat')~_2Y_{\ell m}(\gammahat) + _{-2}Y^*_{\ell' m}(\gammahat')~_{-2}Y_{\ell m}(\gammahat) }{2\gamma(1-\beta  \hat{\bm z} \cdot \gammahat')},
\end{equation}
and 
\begin{equation}
    \Ker{B}{\ell'}{\ell}(\beta)=  \int {\diff^2} \gammahat' \frac{_2Y^*_{\ell' m}(\gammahat')~_2Y_{\ell m}(\gammahat) - _{-2}Y^*_{\ell' m}(\gammahat')~_{-2}Y_{\ell m}(\gammahat) }{-2i\gamma(1-\beta  \hat{\bm z} \cdot \gammahat')}.
\end{equation}
where $_s Y_{\ell m}(\gammahat)$ is the spin weighted spherical harmonics. We should emphasize that these equations are only valid for thermodynamic temperature and do not hold for frequency dependent observables such as brightness temperature $T_\nu$ or specific intensity $I_\nu$. \citet{Yasini:2017jqg} introduced a generalized formalism to deal with the complications arising from the frequency dependence of the motion induced effects. 

\section{TE cross correlation}\label{sec:app:TE}

Throughout the main text we focused on the TT and EE power spectra. In this appendix we discuss how the motion induced effects change the TE power spectrum as well. Since the TE power spectrum goes to zero at certain $\ell$ modes, calculating its relative change $\Delta C^{TE}_\ell/C^{TE}_\ell$ yields infinities over those modes. In order to alleviate this problem in the following plots, instead of dividing the change in the power spectrum $\Delta C^{TE}_{\ell}$ by $C^{TE}_{\ell}$, we divide by the following 

\begin{equation}
    G^{TE}_{\ell} \equiv \sqrt{(C^{TE}_{\ell})^2 + C^{TT}_{\ell} C^{EE}_{\ell}}, 
\end{equation}
which is inspired by the expression for the TE cosmic variance. 

\subsection{Equal-Area Strip Cuts}

\begin{figure}
    \centering
    \includegraphics[width=0.9\linewidth]{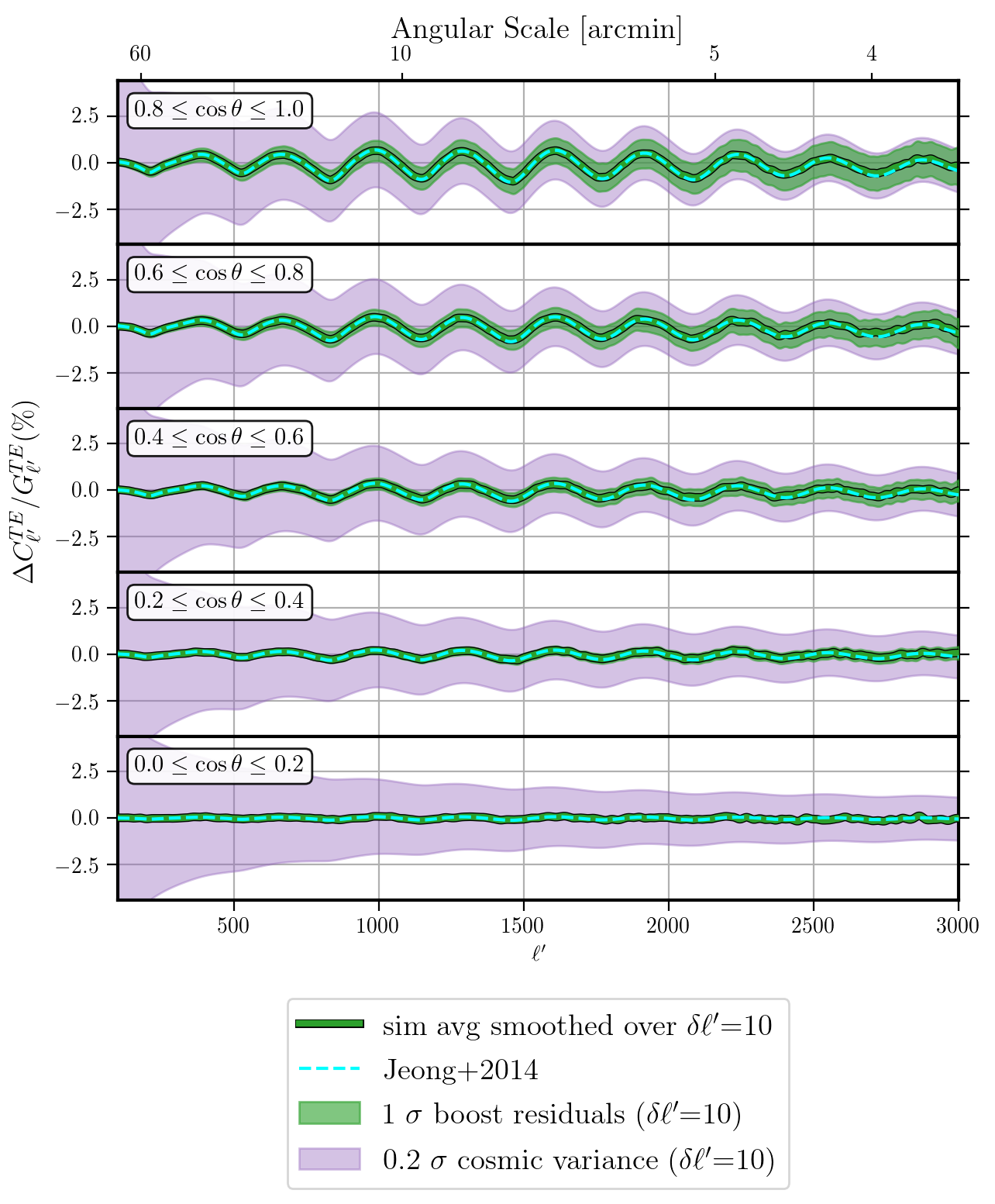}
    \caption{The relative change in the TE power spectrum due to the observer's motion. The amplitude of fluctuations are comparable to TT and EE cases, but unlike them there is no overall increase in power at smaller scales for TE. Similar to TT and EE, the Jeong+14 formula \emph{dashed cyan} emulates the average effect extremely well, but leaves residuals in individual power spectrum realizations (\emph{green band}). The residuals can be as large as 20\% of cosmic variance (\emph{purple band}). } 
    \label{fig:TE_5strip}
\end{figure}

In order to gauge the significance of the motion-induced effects in the TE power spectrum we repeat the exercise of \S~\ref{sec:boost_on_CMB} on the five equal-area strip cuts in the northern hemisphere. Fig. \ref{fig:TE_5strip} shows the relative change in the TE cross correlation of 100 simulations for each of these strips, as well as the Jeong approximation and the 1$\sigma$ region of the residuals it leaves in the power spectrum. Unlike the TT and EE (see Fig. \ref{fig:5strip}) there is no significant increase in the overall power as we go to smaller angular scales. However, similar to TT and EE the amount of fluctuation increases as we go to higher latitudes (closer to the direction of motion). The \citet{Jeong:2013sxy} formula is still performing well for the change in the TE power spectrum, but it still leaves residuals for individual simulations.

\subsection{Hemispherical Asymmetry}

\begin{figure}
    \centering
    \includegraphics[width=0.9\linewidth]{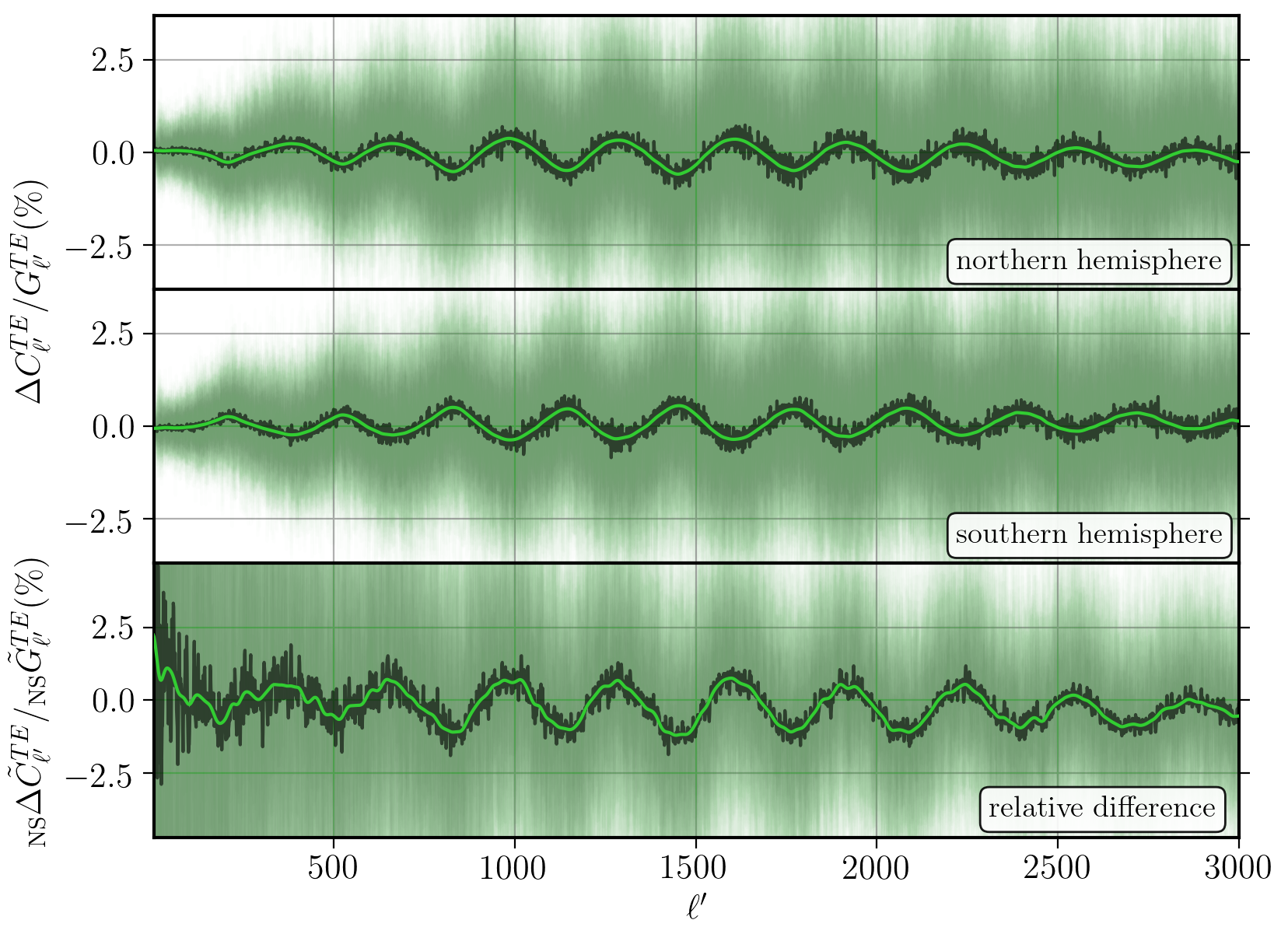}
    \caption{The observer's motion induces an increase in power in the northern hemisphere (\emph{top}), and a decrease in the southern hemisphere (\emph{bottom}), which results in an overall $\sim1-2\%$ hemispherical asymmetry in the TE power spectrum. The \emph{faint green lines} in the background are individual sims, the \emph{jagged black line} is the ensemble average, and the \emph{smooth green line} is the average Gaussian smoothed over $\delta \ell'=10$. The statistic in the bottom panel is the same as the one in  Eq.~\eqref{eq:NS_delta} with $G_{\ell'}$ replacing $C_{\ell'}$ in the denominator. }
    \label{fig:NS_TE_asymmetry}
\end{figure}

Let us look at the amount of hemispherical asymmetry induced in TE due to the motion of the observer. Fig. \ref{fig:NS_TE_asymmetry} shows the amount of induced  asymmetry in the northern and southern hemispheres, as well as the difference between the two. Similar to TT and EE, there is about $\sim1-2\%$ motion-induced hemispherical in the TE power spectrum in the average of 100 simulations. 

\begin{figure}
    \centering
    \includegraphics[width=0.9\linewidth]{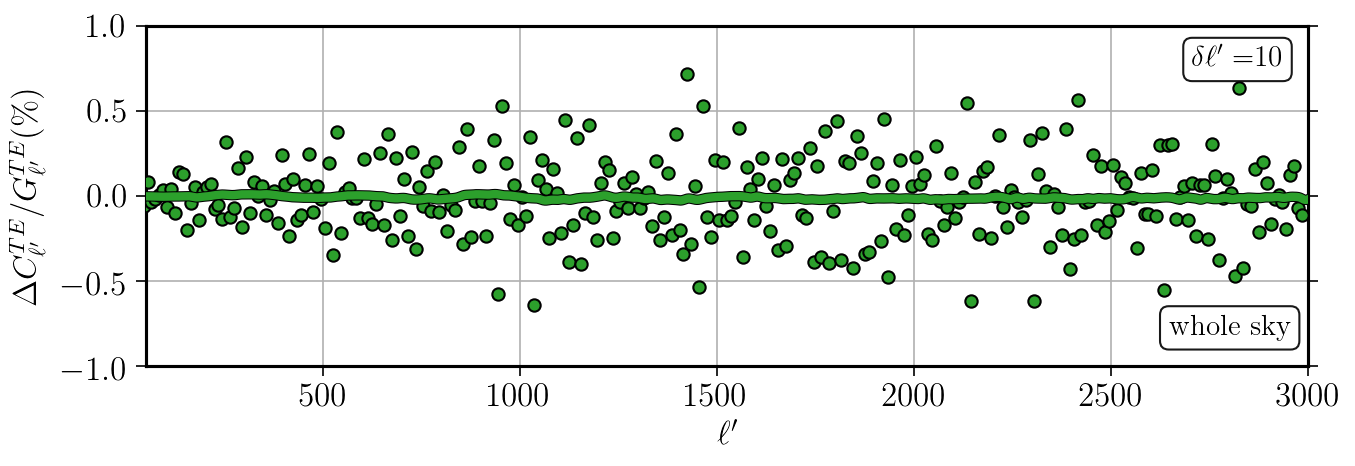}
    \caption{The motion-induced effects on each hemisphere cancel each other out when calculating the power spectrum on the whole sky. }
    \label{fig:TE_fullsky}
\end{figure}

Similar to what happens in the case of TT and EE, the effects on the northern hemisphere cancel out the ones in the southern hemisphere upon calculation of the power over the whole-sky (see Fig.~\ref{fig:TE_fullsky}. The average change in TE drops to about 0.02\% in the whole-sky power spectrum. 

\subsection{Planck SMICA}
\begin{figure}
    \centering
    \includegraphics[width=0.9\linewidth]{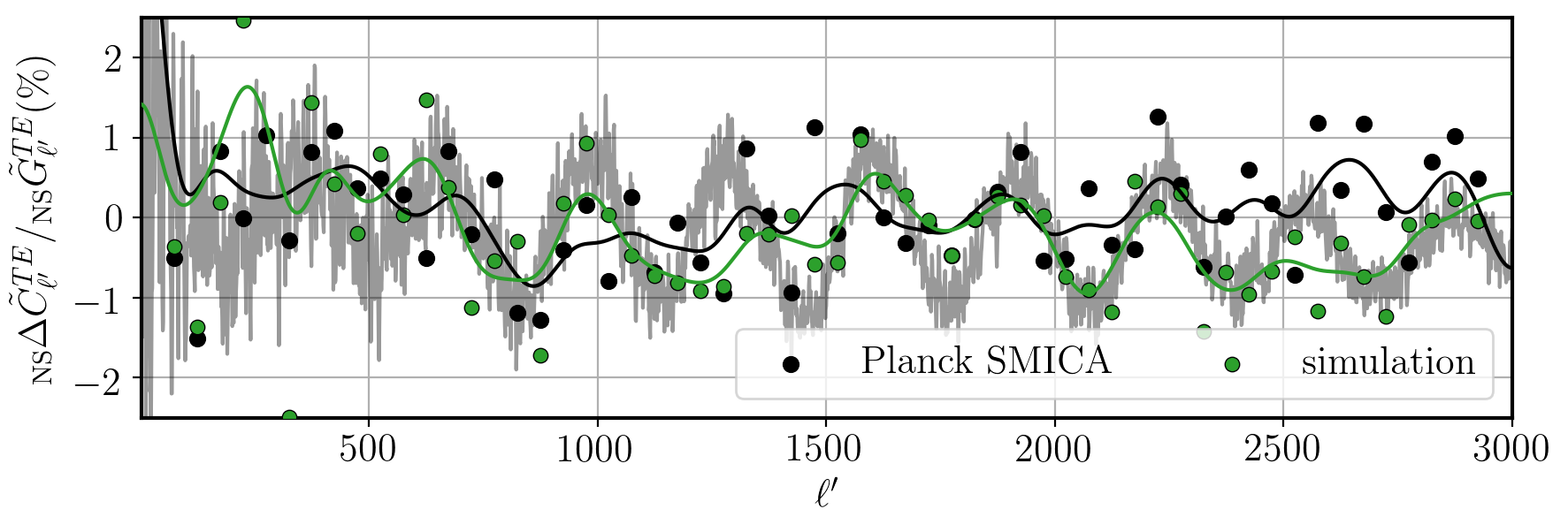}
    \caption{Comparison of Planck SMICA TE hemispherical asymmetry with simulations. The grey lines are the average of 100 simulations, the green circles and  lines are a randomly chosen simulation binned and Gaussian smoothed over $\delta \ell'=50$, and the black circles and lines are the corresponding variables for Planck SMICA. The similarity between the fluctuation pattern of the two lines are certainly suggestive of the Doppler and aberration left-overs in the SMICA TE spectrum. }
    \label{fig:NS_TE_asymmetry_SMICA}
\end{figure}

Finally, in Fig. \ref{fig:NS_TE_asymmetry_SMICA} we look at the the difference between the TE power spectra in the northern and southern hemispheres of the Planck SMICA maps and compare them to the boosted simulations. Similar to the case of TT and EE (Fig. \ref{fig:NS_asymmetry_planck}), the fluctuations in the hemispherical asymmetry of SMICA TE resembles that of the simulations, which suggests that this difference could be at least in part due to the Doppler and aberration effects. We will look at this possibility closely in future work.  

\section{Boost Variance Calculations}
\subsection{Cosmic Variance}\label{sec:app:cosmic_variance}

We start with a quick derivation of the expression for cosmic variance for review. The variance of the spherical harmonic coefficients of an observable such as CMB temperature or polarization is denoted with the theoretical power spectrum $\C_\ell$ as
\begin{equation}
    \langle a^*_{L M} a_{\ell m}\rangle = \delta_{L\ell}\delta_{M m} \C_\ell.
\end{equation}
An unbiased quadratic estimator for the power spectrum can be constructed as
\begin{equation}\label{eq:Cl_estimator}
    C_\ell = \frac{1}{2\ell+1}\sum_m a^*_{\ell m} a_{\ell m},
\end{equation}
which satisfies
\begin{equation}
    \langle C_\ell \rangle = \C_\ell,
\end{equation}
as is expected from an unbiased estimator. Using Eq.~\eqref{eq:Cl_estimator} we can easily calculate the variance of the estimator as 

\begin{align}
    \Delta C_{\ell}^2 =& \langle (C_\ell -\C_\ell)^2\rangle= \langle C_\ell^2\rangle -\C_\ell^2\\
                      =& \langle \frac{1}{(2\ell+1)^2} \sum_{Mm}a^*_{\ell m} a_{\ell m} a^*_{\ell M} a_{\ell M}\rangle -\C_\ell^2\\
                      =& \frac{1}{(2\ell+1)^2} \sum_{Mm}[ \langle a^*_{\ell m} a_{\ell m}\rangle \langle a^*_{\ell M} a_{\ell M}\rangle+\\
                      &\hspace{6em}\langle a^*_{\ell m} a^*_{\ell M}\rangle \langle a_{\ell m} a_{\ell M}\rangle +\\
                      &\hspace{6em}\langle a^*_{\ell m} a_{\ell M} \rangle\langle a^*_{\ell M} a_{\ell m} \rangle ] -\C_\ell^2
\end{align} 
where in the second equality we have used Isserlis' theorem (a.k.a. Wick's theorem). Using the symmetry properties of spherical harmonic coefficients $a^*_{\ell, m}= (-1)^m a_{\ell, -m}$  we can simplify the above equation as 
\begin{align}
    \Delta C_{\ell}^2 =& \frac{1}{(2\ell+1)^2}[(2\ell+1)^2 \C_\ell^2 + 2(2\ell+1)\C^2_\ell] - \C_\ell^2\\
                    =& \frac{2}{(2\ell+1)}\C_\ell^2,
\end{align} 
and hence the well-known formula for cosmic variance 

\begin{equation}\label{eq:cosmic_variance}
    \boxed{\frac{\Delta C_\ell}{\C_\ell} = \sqrt{\frac{2}{2\ell+1} }}.
\end{equation}

\subsection{Boost Variance}\label{sec:app:boost_variance}

In order to simplify the calculation of cosmic  variance in the boosted frame $\langle (\tilde{C_\ell} - \C_\ell)^2\rangle$, we exploit the fact that the labels "rest" and "boosted" frames can be switced under the transformation $\beta \rightarrow -\beta$. In other words\footnote{For simplicity, in this subsection we do not use the prime accent for variables in the boosted frame. }

\begin{equation}
    \tilde{C}_\ell(\beta) - \C_\ell = C_\ell - \tilde{\C}_\ell(-\beta),
\end{equation}
and hence
\begin{equation}\label{eq:equal_variances}
    \langle(\tilde{C}_\ell - \C_\ell)^2\rangle = \langle (C_\ell - \tilde{\C}_\ell)^2\rangle.
\end{equation}
Now we employ the analytical formula of \cite{Jeong:2013sxy} to estimate the motion-induced change in the theoretical power spectrum as 

\begin{equation}
    \tilde{\C}_\ell \simeq \C_\ell - \beta \ell \frac{\diff \C_\ell}{\diff\ell} \overline{\cos\theta} + \frac{1}{2}\beta^2 \ell^2 \overline{\cos^2\theta}\frac{\diff^2 \C_\ell}{\diff\ell^2}.
\end{equation}
Here $\tilde{\C}_\ell$ and $\C_\ell$ respectively represent the power spectra in the boosted and rest frames, and the overline indicates an angular average over the observation patch. After plugging this into Eq. \eqref{eq:equal_variances}, it is straightforward to calculate the variance of the boosted power spectrum around the theoretical background 

\begin{align}
    \langle(\tilde{C}_\ell - \C_\ell)^2\rangle \simeq& \frac{2}{2\ell+1}\C^2_\ell + \left(\beta \ell \overline{\cos\theta}\frac{\diff \C_\ell}{\diff\ell} \right)^2 \nonumber\\ + &\beta^3 \ell^3 \overline{\cos\theta}\overline{\cos^2\theta}\frac{\diff \C_\ell}{\diff\ell} \frac{\diff^2 \C_\ell}{\diff^2\ell} +\mathcal{O}(\beta^4)
\end{align}
The first term is the well-known cosmic variance, and the rest are induced by the motion of the observer. The second term, which we called \emph{boost variance} in the main text, is dominant over the third term and higher order terms which do not appear in the above expression.

\subsection{Moments of the Power Spectrum and its Derivatives}\label{sec:app:moments_of_Cl}

For the sake of completeness we provide another derivation of the expression for cosmic variance using the probability distribution of $C_\ell$. 
We start by considering the distribution function of $C_\ell/\C_\ell$. Assuming that the $a_{\ell m}$ coefficients are Gaussian distributed, the quantity 
\begin{equation}\label{eq:chi_definition}
    \chi_k \equiv (2\ell+1)\frac{C_\ell}{\C_\ell} = \sum_m  \frac{|a_{\ell m}|^2}{\C_\ell},
\end{equation}
has the following chi-squared distribution

\begin{equation}\label{eq:chi2_dist}
    P_k(\chi_k) = \frac{\chi_k^{k/2-1} e^{-\chi_k /2}}{2^{k/2}\Gamma(k/2)} , \hspace{2em} k\equiv 2\ell+1.
\end{equation}
By evaluating the first moments of $\chi_k$ we obtain 

\begin{equation}\label{eq:chi_mean}
    \langle \chi_k \rangle = \int P_k(\chi_k) \chi_k \diff{}\chi_k = k
\end{equation}
which is equivalent to 

\begin{equation}
    \boxed{\langle C_\ell \rangle = \C_\ell}.
\end{equation}
From the second moment 

\begin{equation}\label{eq:chi_var}
    \langle {\chi_k}^2 \rangle = \int P_k(\chi_k) {\chi_k}^2 \diff\chi_k = k^2 + 2k
\end{equation}
we find 
\begin{equation}\label{eq:<Cl^2>}
    \boxed{
    \langle C^2_\ell \rangle = \frac{2 \ell+3}{2\ell+1}\C_\ell^2
    }
\end{equation}
which is equivalent to Eq.~\eqref{eq:cosmic_variance}. 

As for the derivatives of $C_\ell$, we can calculate them by taking the derivative of Eq.~\eqref{eq:chi_mean} with respect to $k$

\begin{equation}\label{eq:dchi_dk}
    \langle \partial_k\chi_k \rangle = \int P_k(\chi_k) \frac{\diff \chi_k }{\diff k}\diff{}\chi_k = 1,
\end{equation}
where $\partial_k \equiv \frac{\diff}{\diff k} = \frac{\diff}{2\diff \ell}$. The caveat here is that the ensemble average and the derivative do not necessarily commute for low values of $k$, but when $k \gg 1$ we have $\int \partial_k P_k \chi_k \diff{} \chi \simeq 0$ so Eq.~\eqref{eq:dchi_dk} is valid. Rewriting this equation in terms of $C_\ell$ we obtain 

\begin{equation}
    \langle \frac{C_\ell}{\C_\ell} + k \frac{\partial_k C_\ell \C_\ell - C_\ell \partial_k \C_\ell }{\C_\ell^2} \rangle = 1,
\end{equation}
which simplifies to the following expression for the ensemble average of the derivative of the power spectrum:
\begin{equation}
    \boxed {
    \langle \frac{\diff C_\ell}{\diff \ell} \rangle = \frac{\diff \C_\ell}{\diff \ell} 
    }.
\end{equation}
The average of the second derivative of $C_\ell$ can be derived similarly as 
\begin{equation}
    \boxed {
    \langle \frac{\diff^2 C_\ell}{\diff \ell^2} \rangle = \frac{\diff^2 \C_\ell}{\diff \ell^2} 
    }.
\end{equation}
One can take this a step further and calculate the cross-correlation between the power spectrum and its derivative. Taking the derivative of Eq.~\eqref{eq:chi_var} yields 

\begin{equation}\label{eq:chi-dchi_dk}
    \langle \chi_k \partial_k\chi_k \rangle = k+ 1.
\end{equation}
After rewriting this in terms of $C_\ell$ we find 

\begin{equation}\label{eq:<CldCl>}
    \boxed {
    \langle C_\ell \frac{\diff C_\ell}{\diff \ell} \rangle = \frac{2\ell+3}{2\ell+1} \C_\ell \frac{\diff \C_\ell}{\diff \ell} - \frac{2}{(2\ell+1)^2} \C^2_\ell 
    }.
\end{equation}
Note that since the derivative and the ensemble average commute, the same exact expression could have been obtained by taking the derivative of \ref{eq:<Cl^2>} with respect to $\ell$. From \ref{eq:<CldCl>} we can also find the first moment of the derivative of $C^2_\ell$

\begin{equation}\label{eq:<dCl^2>}
    \boxed {
    \langle \frac{\diff C^2_\ell}{\diff \ell} \rangle = 2\langle C_\ell \frac{\diff C_\ell}{\diff \ell} \rangle 
    }.
\end{equation}
Higher order statistics for the power spectrum and its derivative can be derived in a similar fashion.

\bsp	
\label{lastpage}
\end{document}